\documentclass[useAMS,usenatbib]{mn2e}
\usepackage{psfig}
\usepackage{amssymb}
\usepackage{color}
\usepackage{enumerate}
\usepackage{rotating}
\usepackage{amsmath}
\usepackage{mathrsfs}
\usepackage{ragged2e}
\usepackage{hyperref}
\usepackage{comment}

\newcommand{\be}{\begin{equation}}
\newcommand{\ee}{\end{equation}}
\newcommand{\bea}{\begin{eqnarray}}
\newcommand{\eea}{\end{eqnarray}}
\newcommand{\no}{\nonumber}

\title[The MTOE applied to VIPERS]
{The Multi-Tracer Optimal Estimator applied to VIPERS}

\author[Montero-Dorta et al.]{
\parbox[t]{\textwidth}{
Antonio D. Montero-Dorta$^{1}$\thanks{E-mail: amonterodorta@gmail.com},  L. Raul Abramo$^{1}$, Benjamin R. Granett$^{2,3}$, \\
Sylvain de la Torre$^{4}$ \& Luigi Guzzo$^{2,3,5}$}
\vspace*{6pt} \\ 
$^1$ Departamento de F\'isica Matem\'atica, Instituto de F\'isica, Universidade de S\~ao Paulo, Rua do Mat\~ao 1371, CEP 05508-090, \\
S\~ao Paulo, Brazil \\
$^2$ Universit\`a degli Studi di Milano, via G. Celoria 16, 20133 Milano, Italy \\
$^3$ INAF - Osservatorio Astronomico di Brera, via Brera 28, 20122 Milano, Italy \\
$^4$ Aix-Marseille Univ, CNRS, CNES, LAM, Marseille, France\\
$^5$ INFN – Sezione di Milano, Via G. Celoria 16, 20133, Milano, Italy \\
\vspace{-0.4cm} 
}

\date{Accepted ---. Received ---;in original form --- \vspace{-0.3cm}}

\def\simlt{\lower.5ex\hbox{$\; \buildrel < \over \sim \;$}}
\def\simgt{\lower.5ex\hbox{$\; \buildrel > \over \sim \;$}}
\usepackage{graphicx}
\usepackage{rotating}

\definecolor{red}{rgb}{1,0,0}

\begin{document}

\bibliographystyle{mnras}

\maketitle

\begin{abstract}

We use mock galaxy data from the VIMOS Public Extragalactic Redshift Survey (VIPERS) to test the performance of the {\it{Multi-Tracer Optimal Estimator}} (MTOE) of Abramo et al. as a tool to measure the monopoles of the power spectra of multiple tracers of the large-scale structure, $P^{(0)}_\alpha(k)$. We show that MTOE provides more accurate measurements than the standard technique of Feldman, Kaiser \& Peacock (FKP), independently of the tracer-selection strategy adopted, on both small and large scales. The largest improvements on individual $P^{(0)}_\alpha(k)$ are obtained using a colour-magnitude selection on small scales, due to MTOE being naturally better equipped to deal with shot noise: we report an average error reduction with respect to FKP of $\sim$ 40$\%$ at $k \, [h$ Mpc$^{-1}]\gtrsim 0.3$. On large scales ($k[h$ Mpc$^{-1}]\lesssim0.1$), the gain in accuracy resulting from cosmic-variance cancellation is $\sim$ 10$\%$ for the ratios of $P^{(0)}_\alpha(k)$. We have carried out a Monte-Carlo Markov Chain analysis to determine the impact of these gains on several quantities derived from $P^{(0)}_\alpha(k)$. If we push the measurement to scales $0.3 < k \, [h$ Mpc$^{-1}]< 0.5$, the average improvements are $\sim$ 30 $\%$ for the amplitudes of the monopoles, $\sim$ 70 $\%$ for the monopole ratios, and $\sim$ 20 $\%$ for the galaxy biases. Our results highlight the potential of MTOE to shed light upon the physics that operate both on large and small cosmological scales. The effect of MTOE on cosmological constraints using VIPERS data will be addressed in a separate paper.

\end{abstract}


\section{Introduction}
\label{sec:intro}

Upcoming dark-energy surveys such as Euclid\footnote{https://www.euclid-ec.org}, the Dark Energy Spectroscopic Instrument (DESI\footnote{https://www.desi.lbl.gov}),
the Dark Energy Survey (DES\footnote{https://www.darkenergysurvey.org}), the Javalambre Physics of the Accelerated Universe Astrophysical Survey (J-PAS\footnote{http://www.j-pas.org}), the  Prime Focus Spectroscopy survey (PFS\footnote{https://pfs.ipmu.jp}),  the Large Synoptic Spectroscopic Telescope (LSST\footnote{https://www.lsst.org}), or SPHEREx\footnote{http://spherex.caltech.edu} will collect data from tens to hundreds of millions of galaxies and quasars over enormous cosmological volumes. These data sets will map the large-scale structure of the Universe (LSS) over vast distances and timescales, allowing highly accurate measurements of the accelerated expansion, the growth rate of structure, and many other cosmological parameters. In addition to their statistical power, some of these upcoming dark-energy surveys are also advantageous, as compared to previous experiments, in that they employ wider galaxy selections that are not based on a few specific LSS tracers -- such as the luminous red galaxies and quasars selected for the Baryon Acoustic Oscillation Spectroscopic Survey (BOSS, \citealt{Eisenstein2011, Dawson2013}).

This upcoming wealth of cosmological-survey data naturally poses the question of how clustering information from different galaxy populations or data sets can be combined in an optimal way. Importantly, different LSS maps are not independent, since different species of galaxies trace the same underlying dark-matter density field, albeit in slightly different ways \citep{Guzzo1997, Benoist1996}. This imposes a constraint between different tracers that can be exploited in order to realize the full potential of overlapping data sets. However, the inclusion of this additional information brings with it the correlations that exist between the distributions of those different tracers, which arise primarily as a consequence of {\it cosmic variance} -- i.e., the fact that local patterns of over- and under-densities propagate to the number counts of all tracers.

Despite this limitation, the strict bounds imposed by cosmic variance do not apply to many key cosmological quantities \citep{Seljak2009,McDonald2008}. By comparing the clustering of different tracers we can {\it{beat cosmic variance}}, i.e., we can measure some quantities with an accuracy that is basically unconstrained by cosmic variance. One of the quantities of interest is {\it{bias}}, which broadly describes the relation between the spatial distribution of galaxies (or haloes) and that of the underlying matter density field. The cancellation of cosmic variance can serve not only to measure bias, but also to improve our constraints on the matter growth rate, reflected in the redshift-space distortions (RSDs, see \citealt{Kaiser1987, Hamilton1998}), as well as the non-gaussianity parameters $f_{NL}$ and $g_{NL}$, and in general any parameters that affect the relative clustering of different tracers (see, e.g., \citealt{Slosar2009,GilMarin2010,Cai2012,Hamaus2011,Smith2011,Hamaus2012,Abramo2013,Blake2013,Ferramacho2014,Bull2015,Fonseca2015,Abramo2017,Alarcon2016,Witzemann2018} for more information).

In this paper, we apply the {\it{Multi-Tracer Optimal Estimator}} (hereafter, MTOE) to mock data that mimic the VIMOS Public Extragalactic Redshift Survey \citep{Guzzo2014} in the redshift range $0.55<z<1.1$. The MTOE for the redshift-space power spectra of an arbitrary number of LSS tracers was derived in \citet{Abramo2016}. The technique, which is a generalization of the standard weighting scheme of \citet*{FKP} (henceforth FKP), is based on the Fisher matrix for multiple tracers \citep{Abramo2012, Abramo2013}, which encapsulates the information contained in multi-tracer cosmological surveys. 

The goal of this paper is to evaluate the performance of MTOE as compared to the traditional FKP approach in the context of the measurement of galaxy power spectra and galaxy biases. Note that FKP does not necessarily imply a single-tracer approach: we will compare MTOE with FKP in a multi-tracer fashion, i.e., in both cases splitting our sample in multiple galaxy populations. The latter approach was employed by \cite{Blake2013}, who analyzed multiple tracers in the Galaxy And Mass Assembly (GAMA) survey. The authors employed FKP weighting in a multi-tracer context, reporting improvements of 10-20 $\%$ in measurements of the gravitational growth rate compared to a single-tracer analysis. The added value of MTOE is that it employs a {\it{multi-tracer weighting scheme}}, which is optimal for measuring the power spectra of the individual tracers.

The first multi-tracer weighting scheme was proposed by Percival, Verde \& Peacock (PVP, \citealt{PVP}). The PVP method provides optimal weights for a minimum-variance estimator of the {\it{combined}} matter power spectrum $P_m(\mathbf{k})$ in situations where several different biased tracers are considered -- see also \citet{Cai2011} and \citet{Granett2015}. As compared to PVP, the advantage of MTOE resides in the fact that it is optimal both in terms of the estimation of the matter power spectrum, as well as the redshift-space auto-power spectra of each individual tracer. In fact, as shown in \citet{Abramo2016}, when the individual power spectra of the tracers estimated with the MTOE method are combined to obtain the matter power spectrum, the result is identical to that obtained using the PVP method.

We have successfully applied MTOE to the determination of the power spectra and the ratios of biases for different subsets of haloes in the context of the study of halo assembly bias (or more generally, {\it{secondary bias}}). In \cite{SatoPolito2018}, we show that MTOE provides a significant improvement in the signal-to-noise of the secondary-bias measurement. In \cite{SatoPolito2018}, each subset of haloes is regarded as a different tracer, and a total of 32 tracers per simulation box are considered. Applying the method to a real galaxy data set is, however,  beset by two difficulties which can potentially reduce the ``gain signal''. First, the natural stochasticity in the way galaxies populate haloes -- i.e., how a particular subset of galaxies might {\it{typically}} occupy certain types of haloes, but only in a statistical sense. Second, the observational limitations associated with real data, such as incompleteness or a tracer-dependent selection function.

VIPERS is an excellent data set to test the performance of MTOE, due to a combination of large volume, broad galaxy selection and high completeness (see, e.g., \citealt{Granett2015,Rota2017,DeLaTorre2017,Pezzotta2017,Mohammad2018} for other cosmological analyses using VIPERS). In order to partially circumvent the second obstacle mentioned above, i.e., the fact that unaccounted observational effects can reduce the potential gain provided by MTOE, in this first paper we concentrate on mock galaxy data. This allows us to build the machinery and to identify the particular cases where MTOE is more advantageous than the standard FKP analysis.   

The paper is organized as follows. In Section~\ref{sec:MTOE}, we summarize the MTOE and FKP techniques by describing their weighting schemes and the treatment of the window functions. Section~\ref{sec:data} describes the VIPERS mock data, and Section~\ref{sec:tracer} presents our tracer selection schemes. Section~\ref{sec:Pk} provides practical information on the implementation of the MTOE method. In Section~\ref{sec:results}, we present the main results of this analysis: the improvements provided by MTOE as compared to FKP on the estimation of the power spectra of multiple tracers and derived quantities such as galaxy biases. Finally, we discuss the implications of our results and summarize our main conclusions in Section~\ref{sec:discussion}. Throughout this work, we assume the standard $\Lambda$CDM cosmology \citep{planck2014}, with parameters $h = 0.677$, $\Omega_m = 0.307$, $\Omega_{\Lambda} = 0.693$, $n_s = 0.96$, and $\sigma_8 = 0.823$.


\section{Optimal estimators of the power spectrum}
\label{sec:MTOE}

\subsection{The optimal weights}

Galaxy surveys are not only limited to finite regions of space, both in the angular and radial direction, they typically have modulations in the mean number densities of galaxies across the survey volume. These variations in the mean density of galaxies (the selection functions, $\bar{n}_\mu$) can be due to an intrinsic redshift-dependent abundance of these objects, but they can also arise as a consequence of the depth of the survey, or because of varying observing conditions across different patches of the sky.

Given that selection functions are in general inhomogeneous, some regions of the survey are expected to carry more weight than others: typically, regions with higher densities of galaxies have better clustering signal compared with regions of low number densities. Since this is ultimately a matter of signal {\it versus} noise, the weights applied to each region in the survey volume must know about the mean density of galaxies in that region, but also about the signal that we are trying to measure -- i.e., the strength of the clustering of the tracers. 

The problem of optimizing the signal with respect to noise appeared with the first galaxy surveys \citep{Davis1982,Hamilton1993}, but it was first solved in a formal sense by Feldman, Kaiser \& Peacock (\citealt{FKP}). 
Under the assumption that the galaxy population constitutes a single-biased tracer of the underlying density field, FKP showed that there is a unique weighting scheme that leads to an optimal (minimum-variance) estimator of the power spectrum. 
In order to construct this estimator, one starts by weighting the density contrast of galaxy counts, $\delta_g  (\mathbf{x})= n_g (\mathbf{x})/\bar{n}_g (\mathbf{x}) - 1$:
\be
\label{eq:ffkp}
f_{FKP}(\mathbf{x}) = w_{FKP}(\mathbf{x}) \delta_g (\mathbf{x}) \; .
\ee
The FKP weights are given by:
\be
\label{eq:wfkp}
w_{FKP}(\mathbf{x}) = 
\frac{ \bar{n}_g(\mathbf{x}) \, b_g}{1 + \bar{n}_g(\mathbf{x}) \, P_g(k_{fid})} \; ,
\ee
where $b_g$ is the fiducial value of the galaxy bias and $P_g(k_{fid}) = b_g^2 \, P_m(k_{fid})$ is the fiducial value of the galaxy power spectrum at some characteristic scale $k_{c}$, with $P_m(k)$ being the matter power spectrum.
The weighted field is then used to construct the power spectrum estimator:
\be
\label{eq:Pfkp}
\hat{P}^{FKP} (\mathbf{k}) = N  
\langle \tilde{f}_{FKP} \tilde{f}_{FKP}^* \rangle_\mathbf{k} - P^{shot}_g \; ,
\ee
where $\langle \ldots \rangle_k = 1/\tilde{V}_\mathbf{k} \int_{\tilde{V}_k} d^3 k / (2\pi)^3 (\ldots) $ is an average over the bandpower (bin) $\mathbf{k}$, $N$ is a normalization, and $P^{shot}_g \sim 1/{\bar{n}}$ is the bias of the estimator (i.e., shot noise).
It can be shown \citep{FKP,TegmarkHamilton1998} that, under some reasonable assumptions, the covariance of the estimator in Equation~\ref{eq:Pfkp} is equal to the inverse of the Fisher matrix of the power spectrum. This means that the FKP power spectrum estimator saturates the Cram\'er-Rao bound and is therefore the best possible estimator for that observable.

However, if there are two or more distinct biased tracers occupying the same survey volume, the signal and the noise for all possible auto- and cross-correlations should now be taken into account, leading to a generalization of the FKP weights. If the goal is to compute the matter power spectrum $P_m(\mathbf{k})$, the generalized weights that lead to a minimum-variance estimator of $P_m (\mathbf{k})$ were first found by \citet{PVP} -- see also \citet{Cai2011} for a minimum-variance estimator of the mass density field. However, if the goal is to compute the redshift-space power spectra of all the individual tracers (which can then be combined to find the matter power spectrum), then the optimal weights are those derived in \citet{Abramo2016}. In what follows, we employ both this latter method, which we call the {\em Multi-Tracer Optimal Estimator}, and the standard method (FKP).

Below we summarize the main results and the method presented in \citet{Abramo2016}. The MTOE weighted fields are given by:
\be
\label{Def:f}
f_\mu (\mathbf{x}) = \sum_\nu w_{\mu\nu} (\mathbf{x})
\, \delta_\nu(\mathbf{x}) \; 
\ee
where $\delta_\mu = n_\mu/\bar{n}_\mu -1$ are the density contrasts of the tracers. The multi-tracer weights are expressed in terms of a fiducial model as:
\be
\label{Def:w}
w_{\mu\nu} (\mathbf{x})
= \left[
\delta_{\mu\nu} - 
\frac{ \bar{n}_\mu (\mathbf{x}) \, P_\mu(k_{c}) }
{1 + {\cal{P}}(\mathbf{x},k_{c})}
\right] \,
\bar{n}_\nu (\mathbf{x}) b_\nu \; ,
\ee
where $\delta_{\mu\nu}$ is the Kronecker delta, and we defined ${\cal{P}}(\mathbf{x},k_{c}) = \sum_\mu \bar{n}_\mu (\mathbf{x}) \, P_\mu(k_{c})$. In this Section, whenever we write $P_\mu(\mathbf{k})$ or just $b_\mu$, it is understood that these are the fiducial power spectra and biases (in real or redshift space), whereas the estimated (measured) power spectra are indicated by $\hat{P}_\mu$. The weights of Equation~\ref{Def:w} are a generalization of the FKP weights for the case of multiple tracers of the LSS: in fact, it is easy to see that the multi-tracer weights reduces to Equation~\ref{eq:wfkp} in the case of a single tracer. 

With the weighted fields of Equation~\ref{Def:f}, we can construct optimal estimators for the individual power spectra in a way similar to Equation~\ref{eq:Pfkp}:
\be
\label{eq:PMTOE}
\hat{P}^{MTOE}_\mu (\mathbf{k}) = \sum_{\nu} N_{\mu\nu} 
\left\langle 
\tilde{f}_\nu \tilde{f}^* + c.c. 
\right\rangle_\mathbf{k} - P^{shot}_\mu \; ,
\ee
where we defined $f = \sum_\mu f_\mu$. 
The ``normalizations'' are given by:
\be
N_{\mu\nu} = \frac{\tilde{V}_\mathbf{k}}{4 \, b_\nu^2} \, F_{\mu\nu} \; ,
\ee
where $F_{\mu\nu}$ is the Fisher matrix for the power spectra, which can be expressed as:
\bea
\nonumber
F_{\mu\nu}(\mathbf{k}) &=& \frac{1}{4} \int_V d^3 x 
\int_{\tilde{V}_\mathbf{k}} \frac{d^3 q}{(2\pi)^3} 
\left[
\delta_{\mu\nu}
\frac{\bar{n}_\mu}{P_\mu} \frac{\cal{P}}{1+{\cal{P}}} \right.
\\ \label{eq:Fish}
& & \left. + \bar{n}_\mu \bar{n}_\nu \frac{1-{\cal{P}}}{(1+{\cal{P}})^2} 
\right]
\; ,
\eea
and the integrand above $\bar{n}_\mu=\bar{n}_\mu(\mathbf{x})$, $P_\mu=P_\mu(\mathbf{q})$ and ${\cal{P}} = {\cal{P}} (\mathbf{x},\mathbf{q})$. 
The Fisher matrix in Equation~\ref{eq:Fish} is in fact the inverse of the covariance matrix for the power spectra of the tracers, $F_{\mu\nu}^{-1} = C_{\mu\nu}$, where the covariance of the power spectra in two Fourier bins $\mathbf{k}_i$ and $\mathbf{k}_j$ is given by
$Cov[ P_\mu(\mathbf{k}_i) , P_\nu(\mathbf{k}_j)] = \delta_{ij} \, C_{\mu\nu}$ \footnote{This holds if the bandpowers $\mathbf{k}_i$ and $\mathbf{k}_j$ are sufficiently wide -- see, e.g., \citep{Abramo2012} for a more detailed discussion.}.

The shot noise (or estimator bias) in Equation~\ref{eq:PMTOE} is given by:
\be
\label{eq:shot}
P^{shot}_\mu = \frac{1}{2} \sum_\nu F_{\mu\nu}^{-1} 
\int_V d^3 x \int_{\tilde{V}_k} \frac{d^3 k}{(2\pi)^3} \frac{\bar{n}_\mu}{1+{\cal{P}}} - P_\mu \; .
\ee
It can be easily checked that in the case of a 
single tracer we obtain:
\be
F_{\mu\nu} \to \frac{V \tilde{V}_\mathbf{k} }{2} \left\langle 
\left( \frac{\bar{n}_g}{1+ \bar{n} P_g} \right)^2 \right\rangle_{\mathbf{x},\mathbf{k}} \; ,
\ee
where $\langle \ldots \rangle_{\mathbf{x},\mathbf{k}}$ means average over the survey volume and over the bandpower. 
This expression can then be used in Equation~\ref{eq:shot},
leading to:
\be
P^{shot}_g \to 
\frac{\left\langle 
\frac{\bar{n}_g}{1+ \bar{n} P_g} \right\rangle_{\mathbf{x},\mathbf{k}}}
{\left\langle
\left( \frac{\bar{n}_g}{1+ \bar{n} P_g} \right)^2 \right\rangle_{\mathbf{x},\mathbf{k}} }
 - P_g \; .
\ee
This is in fact the expression for the FKP estimator bias -- i.e., shot noise. If the selection function is sufficiently homogeneous inside the survey volume $V$, and the power spectrum does not vary appreciably inside the Fourier bin $k$, one can get rid of the averages in the ratio and write $P^{shot}_g \to (1+\bar{n}_g P_g)/\bar{n}_g - P_g = 1/\bar{n}_g$, which is the familiar expression for (Poissonian) shot noise.

\subsection{Estimators v. theory: the window functions}

The power spectra estimated either by Equation~\ref{eq:Pfkp} or Equation~\ref{eq:PMTOE} are based on the Fourier transforms of the weighted fields, not on the density contrasts. Since the weighting is applied in configuration space, the Fourier transforms of the weighted fields are convolutions of the Fourier transforms of the weights and of the density contrasts. This means that the estimated power spectra are themselves convolutions of the putative power spectrum -- i.e., there is mode-coupling.

However, with multiple tracers the window functions couple not only the different modes, but the different tracers as well, so that the estimator is related to the true power spectra by:
\be
\label{eq:conv}
\left \langle \hat{P}^{MTOE}_\mu (\mathbf{k}) \right \rangle
= \int \frac{d^3 q}{(2\pi)^3} \, \sum_\nu W_{\mu\nu} (\mathbf{k},\mathbf{q}) \, P_\nu(\mathbf{q}) \; ,
\ee
where the expectation value on the left-hand side is an ensemble average.
We will show below that these window functions possess two key features: (1) they are normalized; and (2) in the continuum limit, and assuming constant selection functions, the window functions $\sim \delta_D (\mathbf{k}-\mathbf{q})$, meaning that there is no mode-coupling.

\begin{figure*}
\begin{center}
\includegraphics[width=0.96\linewidth]{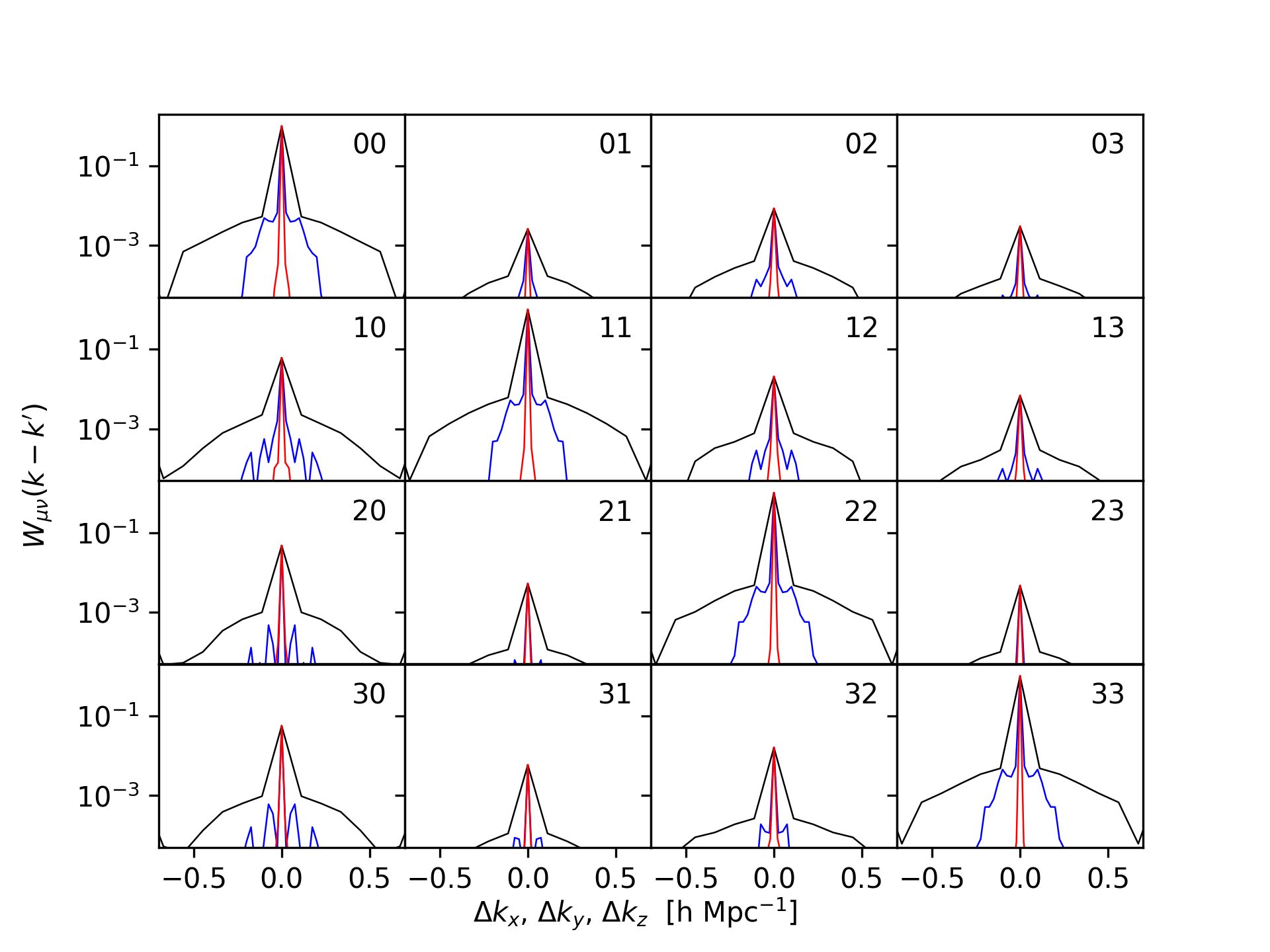}\hfill
\caption{Survey window functions $W_{\mu\nu} (\mathbf{k}-\mathbf{k}')$ for each tracer, as a function of $k_x$ (black), $k_y$ (blue) and $k_z$ (red), normalized at the value of $W_{\mu\mu}(0)$. Rows correspond to $\mu$, columns to $\nu$.}
\label{fig:surveyWF}
\end{center}
\end{figure*}

Although expressions for the window functions were derived in \citet{Abramo2016}, below we expand and clarify those results. It is convenient to write them as:
\be
\label{Eq:WM}
W_{\mu\nu} (\mathbf{k},\mathbf{q}) = \sum_\alpha
\left[ F_{\mu\alpha} (\mathbf{k}) \right]^{-1} M_{\alpha\nu} (\mathbf{k},\mathbf{q}) \; .
\ee
where the matrix $M$ above can be written in terms of the Fourier transforms of the weights,
$\tilde{w}_{\mu\alpha}(\mathbf{q}'-\mathbf{q})$,
as:
\bea
\label{Eq:M1}
M_{\mu\alpha} (\mathbf{k},\mathbf{q})
&=& \frac{1}{4 \, b_\mu^2 \, b_\alpha^2}
\int_{V_\mathbf{k}} 
\frac{d^3 q'}{(2\pi)^3} 
\\ \nonumber
& & \times \sum_{\beta \sigma}
\left[ \,
\tilde{w}_{\mu\alpha} b_\alpha\,
\tilde{w}_{\sigma\beta}^*  b_\beta \, + \,
\tilde{w}_{\mu\beta}  b_\beta\,
\tilde{w}_{\sigma\alpha}^*  b_\alpha\, \right] \; .
\eea
In Figure~\ref{fig:surveyWF}, we show, as an example, the window functions for the VIPERS W1 area within the redshift range $0.6<z<0.75$ (see Section~\ref{sec:data} on the VIPERS data below), assuming a reference $k=0.1$ $h$ Mpc$^{-1}$. Each row $\mu$ describes how a tracer species $\mu$ inherits clustering information from the other tracer species (the columns $\nu$). As can be seen from these plots, the geometry of the survey determines the amount of information in each direction, how that information is mixed, as well as the sampling that can be achieved. 
The largest direction, which yields the cleanest information, is $z$ (which is aligned with redshift), which also gives us the best sampling of frequencies due to the total length of our box in that direction ($L_z \simeq 304 \, h^{-1}$  Mpc in our fiducial $\Lambda$CDM cosmology). 
The $y$ direction (aligned with $RA$, with $L_y \sim 252  \, h^{-1}$ Mpc) also provides good sampling as well as a small mixing with other modes. 
The $x$ direction (aligned with $dec$, with $L_x \sim 58  \, h^{-1}$ Mpc) provides a rather poor sampling ($\Delta k = \pi/L_x \simeq 0.11$ $h$ Mpc$^{-1}$), and is a significant source of mode-mixing. This figure can be compared with Figure 3 of the VIPERS \cite{Rota2017} paper, which shows a similar behaviour regarding the three different directions.

Although Equation~\ref{Eq:M1} is in a form that is convenient to evaluate numerically, in order to check the normalization of the window functions it is useful to Fourier-transform the weights back to configuration space. 
After some algebra we obtain:
\bea 
\nonumber
M_{\alpha\nu} (\mathbf{k},\mathbf{q})
&=& \int_{V_\mathbf{k}} \frac{d^3 q'}{(2\pi)^3} 
\int d^3 x \, d^3 x' \,
e^{i (\mathbf{q}'-\mathbf{q})(\mathbf{x}-\mathbf{x}')}
\\ \nonumber
& & \times \frac{1}{4} \left[ 
\left( \delta_{\mu\alpha} \frac{\bar{n}_\mu}{P_\mu}
- \frac{\bar{n}_\mu \bar{n}_\alpha}{1+{\cal{P}}} \right)_\mathbf{x}
\left( \frac{{\cal{P}}}{1+{\cal{P}}} \right)_{\mathbf{x}'} \right.
\\ \label{Eq:M2}
& & \; + \left.
\left( \frac{\bar{n}_\mu}{1+{\cal{P}}}
\right)_\mathbf{x}
\left( \frac{\bar{n}_\alpha}{1+{\cal{P}}}
\right)_{\mathbf{x}'} \right] \; ,
\eea
where the spectra in the integrand are evaluated at the mean value of the bandpower $\mathbf{k}$.

Now we proceed to demonstrate point (1) above, namely, that the window functions are normalized. Integrating Equation~\ref{Eq:M2} over $\mathbf{q}$ leads to a Dirac delta-function $\delta_D (\mathbf{x} - \mathbf{x}')$, which removes one of the space integrals (say, that on $\mathbf{x}$) and makes the terms inside brackets to coincide at the same point (say, $\mathbf{x}'$). In that case, inspection of Equation~\ref{eq:Fish} immediately reveals that:
\be
\label{Eq:norm1}
\int \frac{d^3 q}{(2\pi)^3} 
M_{\alpha\nu} (\mathbf{k},\mathbf{q}) =
F_{\alpha\nu} (\mathbf{k}) \; ,
\ee
which then implies that:
\be
\label{Eq:norm2}
\int \frac{d^3 q}{(2\pi)^3} 
W_{\mu\nu} (\mathbf{k},\mathbf{q}) =
\delta_{\mu\nu} \; .
\ee
Equation~\ref{Eq:norm2} means that the window functions are not only normalized to unity in terms of the modes, but that the mean mixing between the different tracers vanishes when averaged over all the modes.

Now we turn to the second property of the window function: that in the limit of a large, homogeneous survey, they become delta-functions. The starting point is again Equation~\ref{Eq:M2}, where we now assume that all quantities in the integrand are constant over the survey volume. Since that constant matrix can be taken outside of the spatial integrals, we can perform the $\mathbf{x}$ integral, which leads to
$(2\pi)^3 \delta_D (\mathbf{q} - \mathbf{q}') $.
We also take the bin $\mathbf{k}$ to be arbitrarily small, and the remaining space integral will give the survey volume. Clearly, in that limit we can write that:
\be
\label{Eq:norm3}
M_{\alpha\nu} (\mathbf{k},\mathbf{q}) \to 
F_{\alpha\nu} (\mathbf{k}) \times 
(2\pi)^3 \delta_D (\mathbf{k} - \mathbf{q}) \; ,
\ee
which implies that:
\be
\label{Eq:norm4}
W_{\mu\nu} (\mathbf{k},\mathbf{q}) \to
\delta_{\mu\nu} 
(2\pi)^3 \delta_D (\mathbf{k} - \mathbf{q}) \; .
\ee
This shows that the window functions defined in Equation~\ref{Eq:M1} obey the proper normalization conditions and continuum limit.

In addition to the survey window function, the mass assignment (MA) method, whereby an object (a galaxy, in our case) is associated to a given cell in the grid, may also distort the original power spectrum. In order to deconvolve the effects of the MA on the estimated power spectra, we employ the procedure outlined in \citet{Jing2005}. For all modes of interest, which lie well below the Nyquist frequency of the grid, the MA convolution is well approximated by a normalization factor, i.e., $P(\mathbf{k}) \simeq P(\mathbf{k}) \times W^2_{MA}(\mathbf{k})$. We compute the window function following the standard prescription \citep{Jing2005}, using a theoretical matter power spectrum from CAMB \footnote{\url{https://camb.info}} \citep{CAMB} to account for the shape of the power spectrum.

Finally, we allow for anisotropies in the power spectra through redshift-space distortions. Even though in this first paper we only employ the angle-averaged spectra (i.e., the monopoles), we have also computed the higher-order multipoles. We follow the method proposed by \citet{Bianchi2015} and \citet{Scoccimarro2015}, which makes use of the physical (configuration-space) multipoles of the galaxy maps and avoids having to fix a single direction to the survey in order to define the angles of the Fourier modes ($\mu \to \hat{\mathbf{k}} \cdot \hat{\mathbf{r}}$).

\section{The mock data}
\label{sec:data}

\begin{figure}
\begin{center}
\includegraphics[width=0.96\linewidth]{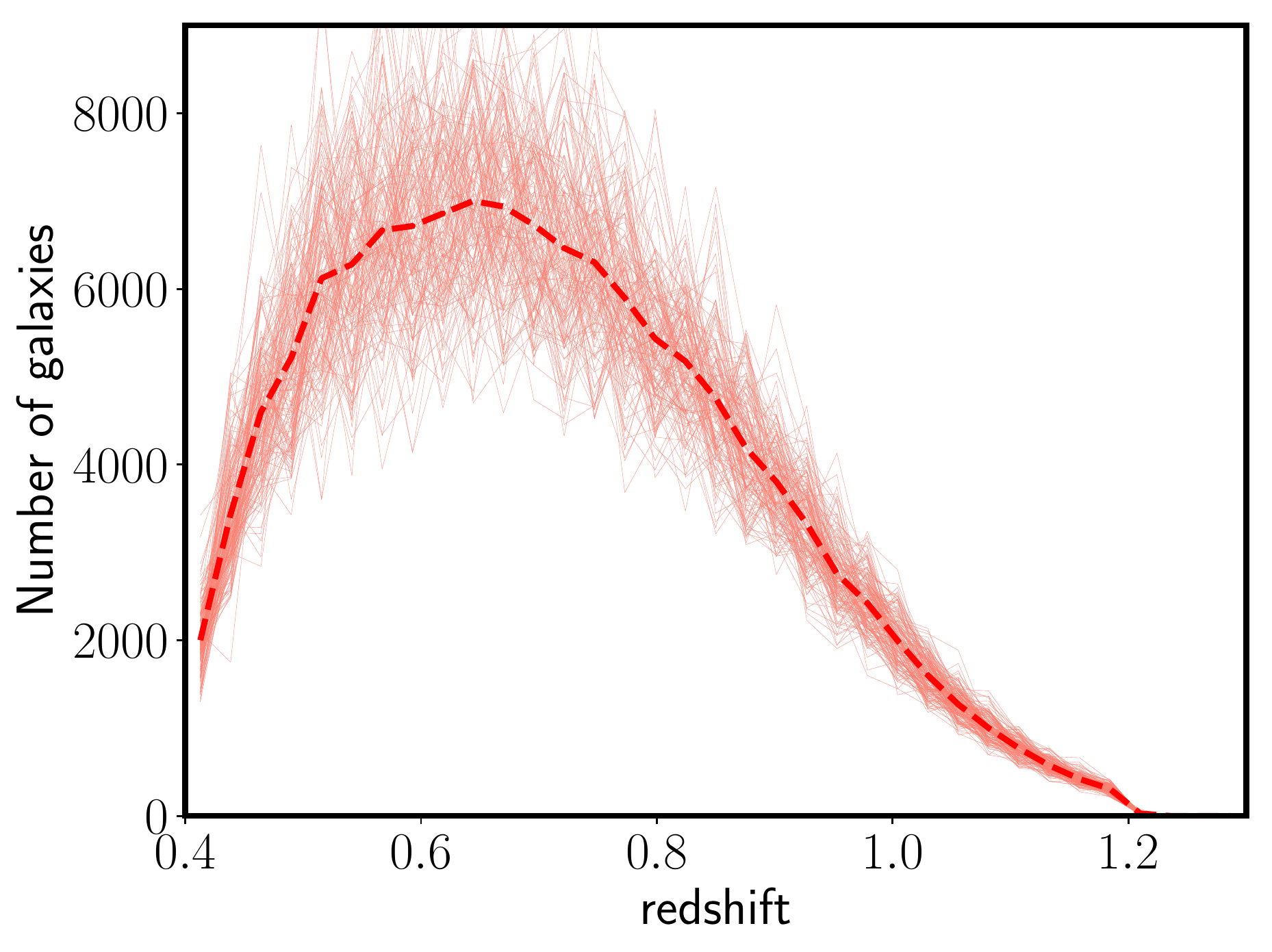}\hfill
\caption{Redshift distribution of the 153 VIPERS mocks used in this work in bins of $\Delta z = 0.025$. The mean 
of all mocks is represented by a thick dashed line.}
\label{fig:zmocks}
\end{center}
\end{figure}

In order to test the performance of MTOE, we use mock data from the VIMOS Public Extragalactic Redshift Survey (VIPERS, \citealt{Guzzo2014,Scodeggio2018}), which provides high-fidelity galaxy maps at high redshift. The survey measured 90,000 galaxies with moderate-resolution spectroscopy within the nominal redshift range $0.5\lesssim z\lesssim1.1$, using the Visible Multi-Object Spectrograph (VIMOS) at VLT.  Targets were selected to a limiting magnitude of $i_{AB}=22.5$ in an area of 24 square degrees of the CFHTLS Wide imaging survey. This area is divided in two fields, W1 and W4, with 16 and 8 square degrees, respectively. The low-redshift limit of the survey was imposed using a colour-based pre-selection that effectively removed foreground galaxies while ensuring a robust flux-limited selection above redshift $z=0.5$. For more information on the Public Data Release 2 (PDR-2), the one used in this paper, refer to \cite{Scodeggio2018}. Detailed descriptions of the data reduction and management infrastructure are presented in \cite{Garilli2014}. For information on the survey design and target selection, see \cite{Guzzo2014}.

We employ a set of 153 independent VIPERS-like mock catalogues \citep{DeLaTorre2013V,DeLaTorre2017,Mohammad2018} that 
mimic the VIPERS survey geometry and the application of the slit-assignment algorithm and redshift measurement error. As discussed in \cite{Mohammad2018}, these mocks represent very well the global properties of the galaxy population targeted by VIPERS. However, they were not built to simultaneously and precisely reproduce the measured bias and number density of multiple sub-populations defined in colour-magnitude space. In order to avoid inconsistencies in the computation of errors, we restrict our analysis to the mocks themselves. In this way, we can better isolate the improvements provided by MTOE and build the machinery for the application of the method to the actual VIPERS data. 

The VIPERS mocks were built from the Big MultiDark Planck (\citealt{Klypin2016}) N-body cosmological simulation, which assumes the standard $\Lambda$CDM cosmology \citep{planck2014}, with parameters $h = 0.677$, $\Omega_m = 0.307$, $\Omega_{\Lambda} = 0.693$, $n_s = 0.96$, and $\sigma_8 = 0.823$. Dark-matter haloes were populated with galaxies using the technique of halo occupation distributions (HOD). The parameters of the HOD were calibrated using clustering measurements as a function of luminosity from the preliminary data release of VIPERS (see \citealt{DeLaTorre2013V, DeLaTorre2017} for a detailed description of the procedure). Importantly, since the resolution of the box is not good enough to match the typical halo masses probed by VIPERS, a population of low mass haloes were subsequently added following the method presented in \citealt{DeLaTorre2013}. In essence, central galaxies were placed at the centre of haloes assuming no peculiar velocities with respect to the hosting-halo rest frame, whereas satellite galaxies were distributed around haloes following an NFW profile \citep{NFW}. For satellite galaxies, a random velocity component was added to the hosting-halo peculiar velocity.
 
Luminosities in the B band and colours were assigned to galaxies following the methodology presented in \cite{Skibba2006} and \cite{Skibba2009}. First, an analytical luminosity- and redshift-dependent HOD parametrization was derived by fitting the observed projected correlation functions in several samples defined using luminosity thresholds \citep{DeLaTorre2013V}. Second, galaxies were placed in haloes following the observed conditional colour distribution ($<U-V|M_B>$) of VIPERS, which was fit with a double Gaussian model. A detailed description on how these assignments were performed can be found in \cite{Skibba2006} and \cite{Skibba2009}. A summary of the procedure can be found in \cite{Mohammad2018}. The redshift distribution of the mocks is shown in Figure~\ref{fig:zmocks}.

\begin{figure}
\begin{center}
\includegraphics[width=0.99\linewidth]{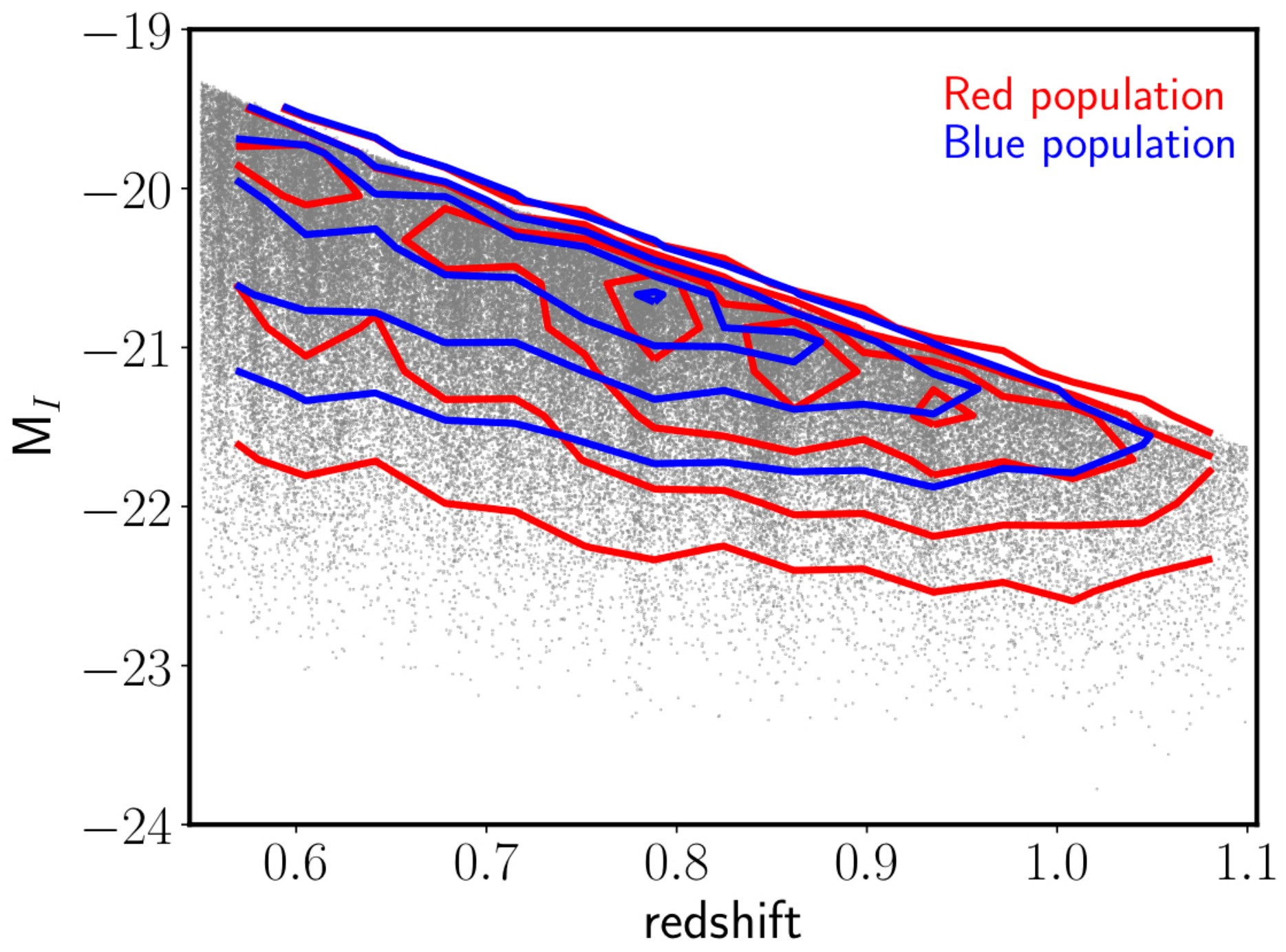}\hfill
\caption{The I-band absolute magnitude as a function of redshift for a randomly-selected mock. The red and blue mock galaxy distributions are represented by red and blue contours, respectively.}
\label{fig:mags}
\end{center}
\end{figure}

\section{Tracer selection}
\label{sec:tracer}

\begin{table*}      
\caption{Number densities $\bar{n}$ in units of [Mpc/h]$^{-3}$ for the 3 tracer selection schemes considered in this analysis. As a reference, the shot-noise level for each subset is typically $\sim 1/\bar{n}$.}  
\begin{center}
\label{tab:schemes}   
\begin{tabular}{ccccccc}        
\hline  Selection scheme &  Criteria  &  Redshift range & $\bar{n}_{T1}$($\times 10^{-4}$) & $\bar{n}_{T2}$($\times 10^{-4}$) &  $\bar{n}_{T3}$($\times 10^{-4}$) & $\bar{n}_{T4}$($\times 10^{-4}$)\\
\hline  L0 &  $M_I$ & $0.55<z<0.7$& 36.6&  46.6& --& --\\
\hline  L &  $M_I$ & $0.55<z<0.7$ &  24.3& 27.7& 19.9& 11.4\\
\hline  LC & $M_I$ \& (U-B) colour   & $0.55<z<0.7$ &  31& 5.6& 36.8& 9.8  \\
   &   & $0.7<z<0.9$ &  17& 3.5& 20.4& 7.6\\
   &   & $0.9<z<1.1$ &  5.2& 1.7& 6.7& 4\\
\hline    
\end{tabular}
\end{center}
\end{table*}

The selection of the LSS tracers is a key part of the multi-tracer analysis, since the performance of the method ultimately depends on our ability to discriminate between different galaxy populations with different biases.

Here we define several selection schemes in order to test the performance of MTOE in different plausible configurations where the number densities and biases of the tracers differ. Our fiducial selection scheme is based on colour and luminosity, which, in principle, should allow for a good separation of tracers in terms of their clustering properties. A total of 4 tracers in this scheme, which we call {\it{LC}} ({\it{luminosity--colour}}), are defined, as follows, in each redshift:\\

\noindent

\noindent Redshift slice $0.55 < z < 0.7$:
\begin{itemize}
	\item[T$_1$:] blue colour \& $M_I > -20.3$ 
		  
	\item[T$_2$:] red colour \& $M_I > -20.3$ 
	
	\item[T$_3$:] blue colour \& $M_I <-20.3$ 
			
	\item[T$_4$:] red colour \& $M_I < -20.3$ 
\end{itemize}

\noindent Redshift slice $0.7 < z < 0.9$:
\begin{itemize}
	\item[T$_1$:] blue colour \& $M_I > -21$ 
		  
	\item[T$_2$:] red colour \& $M_I > -21$ 
	
	\item[T$_3$:] blue colour \& $M_I <-21$ 
			
	\item[T$_4$:] red colour \& $M_I < -21$ 
\end{itemize}

\noindent Redshift slice $0.9 < z < 1.1$:
\begin{itemize}
	\item[T$_1$:] blue colour  \& $M_I > -21.5$ 
		  
	\item[T$_2$:] red colour \& $M_I > -21.5$ 
	
	\item[T$_3$:] blue colour \& $M_I <-21.5$ 
			
	\item[T$_4$:] red colour \& $M_I < -21.5$ 
\end{itemize}

In the VIPERS mocks, a colour type (0:``red'', 1:``blue") is defined from the (U-B) colour, which is in turn computed by matching the colour bimodality of the data (see \citealt{DeLaTorre2013V}). In Figure~\ref{fig:mags}, I-band absolute magnitudes are shown as a function of redshift for a randomly chosen mock, in order to illustrate the LC cuts adopted. The abundances of red and blue galaxies are represented by contours; blue mock galaxies are typically less luminous than red galaxies, at fixed redshift. The LC selection employs an absolute-magnitude cut at the peak of the absolute magnitude distribution at the corresponding redshift.  

It is interesting to analyze how the performance of MTOE depends on the complexity of the selection scheme. To this end, we define a simple 4-tracer scheme where selection is performed by splitting the sample only by luminosity (the {\it{L selection}}). For the sake of simplicity, only the first redshift slice is used for this test. In this case, the 4 tracers are defined as: \\

\noindent Redshift slice $0.55 < z < 0.7$:
\begin{itemize}
	\item[T$_1$ :] $M_I > -20.3$ 
		  
	\item[T$_2$ :]  $-20.3 < M_I < -20.8$
	
	\item[T$_3$ :] $-20.8 < M_I < -21.4$
			
	\item[T$_4$ :] $M_I < -21.4$ 		
\end{itemize}

Finally, we will also consider the most basic selection criterion, i.e., only 2 tracers defined by luminosity:

\noindent Redshift slice $0.55 < z < 0.7$:
\begin{itemize}
	\item[T$_1$ :] $M_I > -20.5$ 
			
	\item[T$_2$ :] $M_I < -20.5$ 		
\end{itemize}

We call this scheme {\it{L0 selection}}. The resulting number density for different tracers in each redshift slice after applying the above selection schemes are listed in Table~\ref{tab:schemes}.

\section{Implementation of the method}
\label{sec:Pk}

In this section, we provide some practical information regarding the main steps in the implementation of the power spectrum estimation methods.

\subsection{Formatting the data}
\label{sec:format}

\begin{figure}
\begin{center}
\includegraphics[width=0.99\linewidth]{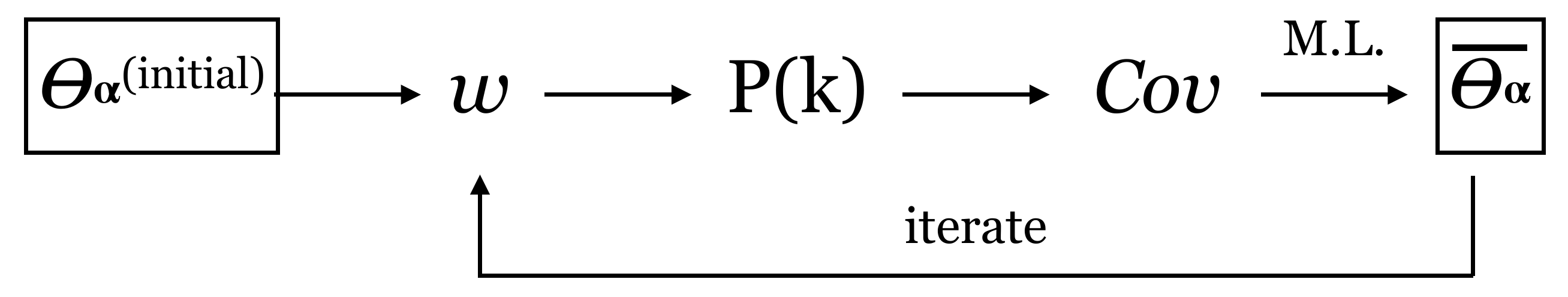}\hfill
\caption{A simple diagram describing the iterative procedure for the estimation of the power spectra. In the first step, the MTOE weights are computed given an initial set of parameters $\theta_{\alpha}^{(initial)}$ (i.e., the galaxy biases, matter power spectrum and matter growth rate). We then use these weights to compute the monopoles of the power spectra of each tracer for each mock. The monopoles of the mocks are then used to determine their covariance matrix. With that covariance matrix we are able, by means of maximum likelihood techniques (i.e. MCMC or $\chi^2$ minimization), to find updated
best-fitting values for the parameters, which can then be used to re-compute the weights and start the process over again. The process is iterated until it converges.}
\label{fig:diagram}
\end{center}
\end{figure}

The practical computation of the power spectrum estimators of Equation~\ref{eq:Pfkp} (FKP) and~\ref{eq:PMTOE} (MTOE) involves binning the galaxy coordinates on a Cartesian grid and using a Fast Fourier Transform (FFT) algorithm. To implement the Fourier-transform approach, galaxies in each of the three redshift slices considered are placed onto three separate grids, of 63$\times$14$\times$76, 73$\times$16$\times$70, and 84$\times$18$\times$84 cells, respectively, with grid cells of $H =$ 4 Mpc$/h$ on a side. The total volumes of the cuboids are 4.29, 5.23, and 8.13 $\times10^{-3}$ [Gpc$/h]^{-3}$, respectively. These dimensions have been chosen so that the maximum fraction of the VIPERS volume is taken into account. We will consider wavenumbers up to $k = 0.5$ $h$ Mpc$^{-1}$ -- well below the Nyquist frequency of the system, which is $\pi/H \sim$ 0.8 $h$ Mpc$^{-1}$.

A cell size of 4 Mpc$/h$ captures the main small-scale features of galaxy clustering and avoids potential issues resulting from very low tracer counts per cell. We have checked that the main conclusions of the analysis remain unaltered when 8 Mpc/h cells are employed.  We note that a smaller 2 Mpc/$h$ cell was used before in the context of the VIPERS power spectrum measurement \citep{Rota2017}, but only for the combined sample.

\subsection{Selection function}
\label{sec:selfunc}

The matter power spectrum is defined through the expectation value (or, in this context, the ensemble average) $ \langle \delta_m (\mathbf{k},z) \delta_m^*(\mathbf{k}',z) \rangle = (2\pi)^3 P_m(\mathbf{k},z) \delta_D (\mathbf{k}-\mathbf{k}')$, where $\delta_m(\mathbf{k})$ is the matter density contrast, and $\delta_D$ is the Dirac delta function. 
For a given tracer with counts per unit volume $n_\alpha(\mathbf{x})$, the density contrast is $\delta_\alpha (\mathbf{x}) = n_\alpha(\mathbf{x})/\bar{n}_\alpha(\mathbf{x}) -1$. Here, the mean number densities $\bar{n}_\alpha$ should reflect the spatial modulations in the observed numbers of galaxies which are due to the instrument, the strategy and schedule of observations, as well as any other factors unrelated to the redshift-space cosmological fluctuations. For simplicity, instead of using random maps or a selection function defined on sub-grid scales, we compute the selection function of each tracer, $\bar{n}_\alpha(\vec{x})$, by taking the mean of all mocks. 
The approximation implied by this procedure can affect the small-scale clustering of sparse samples when the number of mocks is not large enough (i.e., the selection function can be 0 in cells that belong to unmasked survey regions). We have checked that this is not an issue given the size of our cells and the number of mocks available.

\begin{figure*}
\includegraphics[width=1\linewidth]{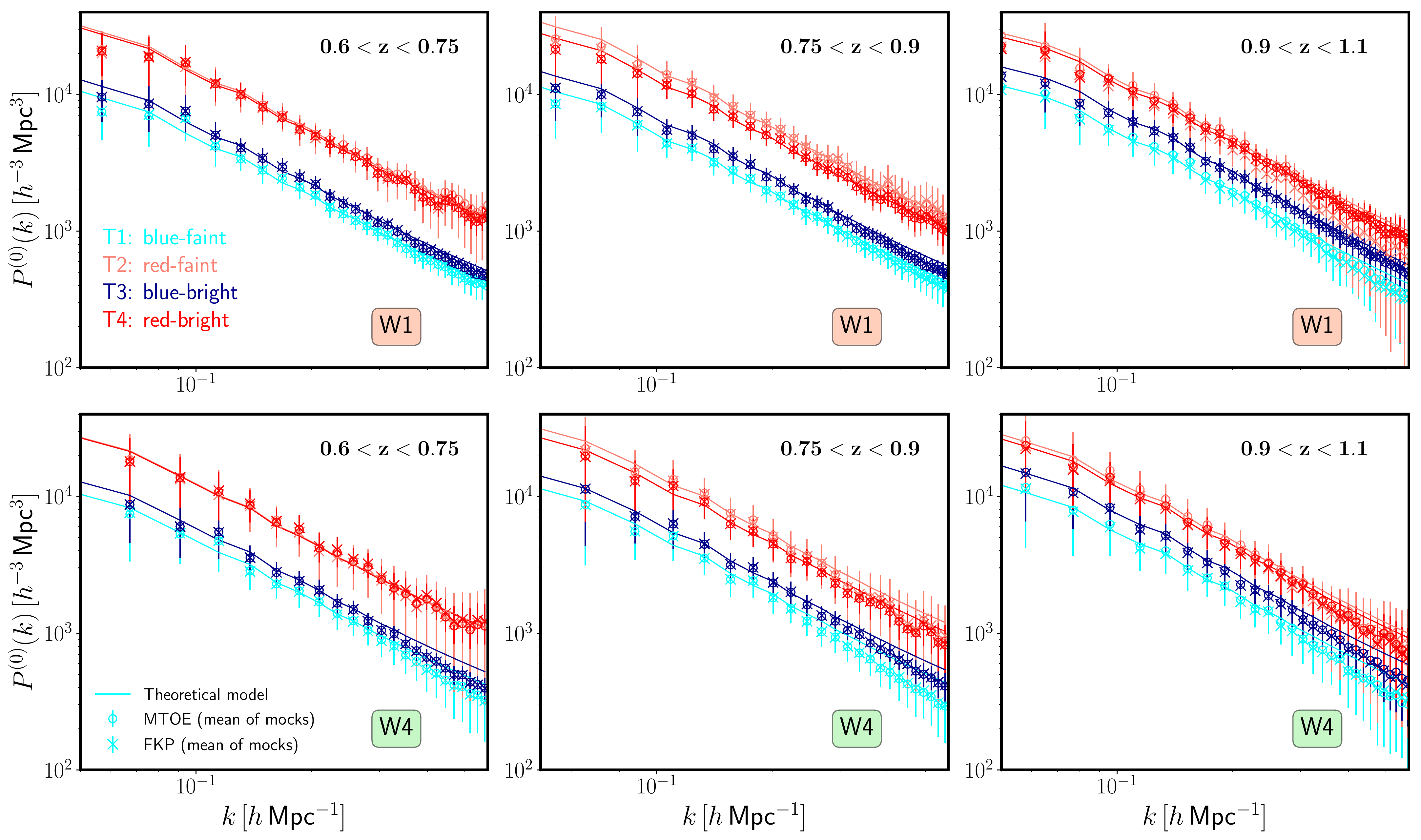}\hfill
\caption{The window-function-deconvolved monopoles of the power spectrum, $P^{(0)}(k)$, for each of the 4 tracers of the LC (luminosity and colour) selection in the three redshift slices considered (for both the W1 and the smaller W4 field). In each panel, the MTOE measurement is shown in circles, whereas the FKP estimate is represented by crosses. For each tracer, a theoretical model assuming the estimated bias value is shown in solid lines (see text).} 
\label{fig:Pk}
\end{figure*}

\subsection{Normalization of the power spectrum}
\label{sec:normalization}

As described in Section~\ref{sec:MTOE}, the ``bare'' estimated power spectra are related to the true power spectra by the convolution expressed in Equation~\ref{eq:conv}. Although for VIPERS this convolution does not significantly distort the shape or the amplitude of the original power spectra, in order to present the data we have normalized the estimated power spectra by the convolved theoretical spectra, i.e.:
\be
\label{Eq:normP}
\bar{P}_\mu (\mathbf{k}) = \frac{1}{N^c_\mu(\mathbf{k})} \, \hat{P}_\mu  (\mathbf{k}) \, ,
\ee
where the normalization is given by the convolution of Equation~\ref{eq:conv} with a theory (CAMB) power spectrum:
\be
N^c_\mu (\mathbf{k}) = \frac{1}{P^{Th}_\mu(\mathbf{k})} \sum_\nu \int \frac{d^3 k'}{(2\pi)^3} W_{\mu\nu} P^{th}_\nu (\mathbf{k}')
\ee
In practice, this normalization is computed independently of the data/mocks using log-normal multi-tracer maps, which are produced using a generalization of the procedure of \citet{Coles1991} (see also \citealt{Abramo2016}). 
These sets of maps are created for each tracer using the same selection functions and formatting that were described in Section~\ref{sec:format} and Section~\ref{sec:selfunc} .

\subsection{Fiducial model}
\label{sec:theory}

Throughout this paper, we use a theoretical model for the 
matter power spectrum $P_m(k)$ obtained from the Boltzmann code \texttt{CAMB}\footnote{http://CAMB.info} \citep{CAMB}, at the corresponding median redshift of the particular redshift slice considered, and assuming the Planck cosmology. 

For the galaxy power spectrum of a tracer, the CAMB model is combined with HaloFit \citep{HaloFit}, which implements non-linear corrections in the clustering. We have checked that our results, in terms of the FKP v. MTOE comparison, remain unaltered when a linear power spectrum is employed.

Fiducial values for the galaxy biases of the sub-populations are obtained iteratively. To initiate the process, we use values that roughly follow the magnitude and colour dependence of bias that was previously measured from VIPERS (see Figure 8 from \citealt{Granett2015}). These biases are used to generate the MTOE weights that are subsequently applied to the density field in order to compute $P(k)$. This computation is performed for all mocks, which allows us to obtain the covariance matrix. From the covariance matrix we determine best-fitting biases by means of a likelihood maximization technique (such as $\chi^2$ minimization or MCMC). These updated biases are injected back into Equation~\ref{Def:w} in order to compute the MTOE weights, and the procedure is iterated until convergence. See Figure~\ref{fig:diagram} for a simple diagram describing the process.

We have checked that the convergence is rather fast (a few iterations suffice), and that the procedure is robust, even in cases where the initial fiducial values (the ones that we used to initialize the process) are completely off (e.g. all biases set to 1). As a reference, for the first redshift slice and the W1 field (LC selection), if we initialize the process with all bias values equal to unity, the procedure takes 3 iterations to converge at a 2$\%$ level. Note that this convergence level is significantly below the final errors that we obtain for the biases.

\section{Results}
\label{sec:results}

In this section, we evaluate the performance of MTOE as compared to FKP in the context of the determination of the power spectra (the basic measurement) and the galaxy biases of multiple tracers. The posterior probability density function of the galaxy biases and related quantities is obtained using a Monte Carlo Markov Chain (MCMC) procedure. 

\subsection{The power spectrum}
\label{sec:results_pk}

Figure~\ref{fig:Pk} presents the monopoles of the normalized power spectra (see Equation~\ref{Eq:normP}) of the four tracers of the LC selection, in each of the 3 redshift bins considered, i.e., $0.6<z<0.75$, $0.75<z<0.9$, and $0.9<z<1.1$, respectively, for both the W1 and the W4 fields. The symbols correspond to the means of the mocks, and the error bars are obtained from the diagonal of the (sample) covariance matrix. In each panel, both the FKP and the MTOE estimates are shown for comparison, along with the fiducial theoretical model (assuming the best-fitting bias values obtained through the MCMC procedure, see Section~\ref{sec:results_mcmc} below). One of our nuisance parameters is a constant additional shot noise term, which only makes a (small) difference in the case of very sparse tracers, and on small scales.

\begin{figure*}
\begin{center}
\includegraphics[width=0.9\linewidth]{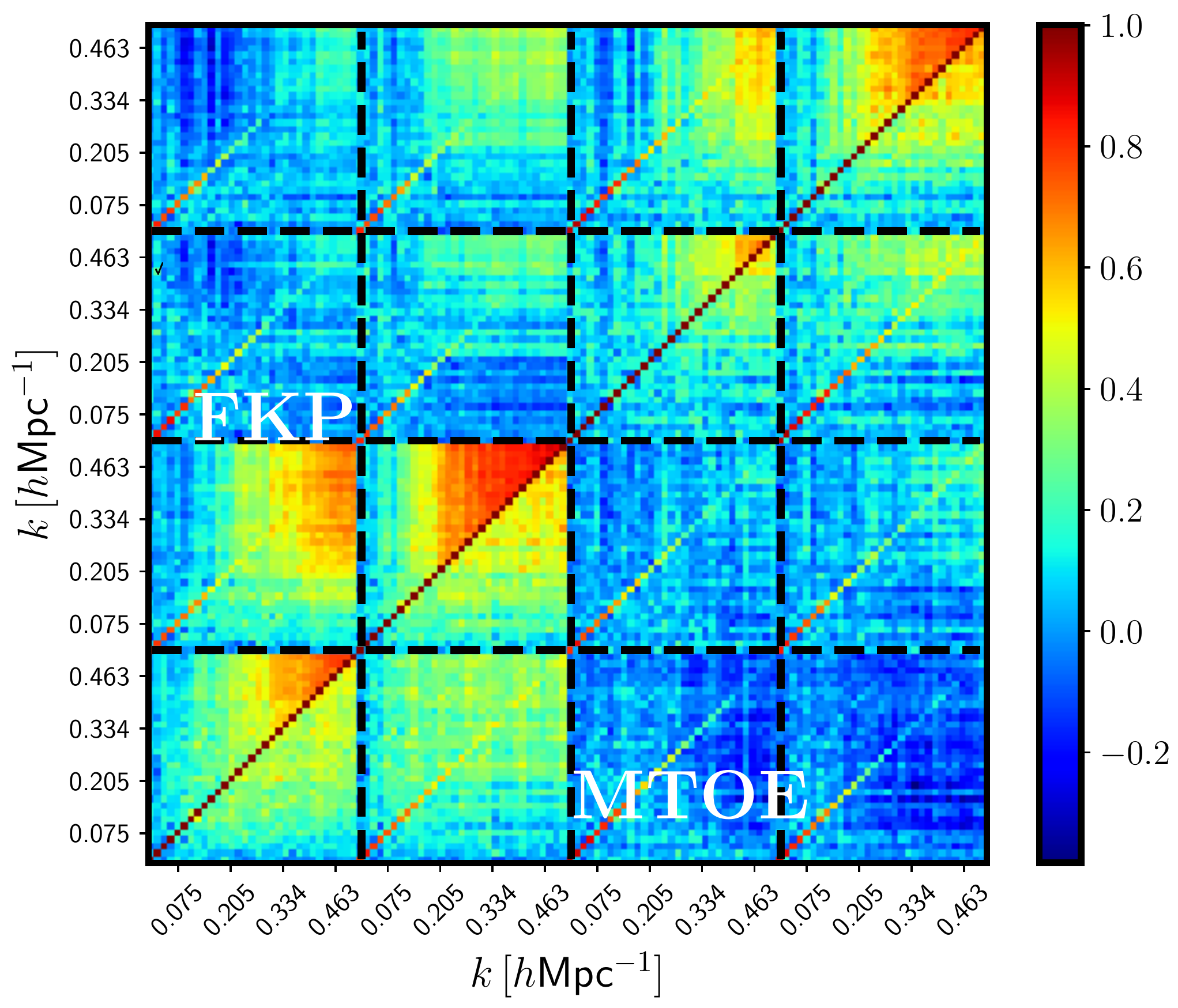}\hfill
\caption{The complete correlation matrix for the entire set of tracers of the LC selection in the first redshift slice, i.e., $0.6<z<0.75$ (W1 field). The MTOE method consistently provides a less-correlated measurement than FKP, especially on small scales.}
\label{fig:corrmatrix_total}
\end{center}
\end{figure*}

\begin{figure*}
\begin{center}
\includegraphics[width=0.9\linewidth]{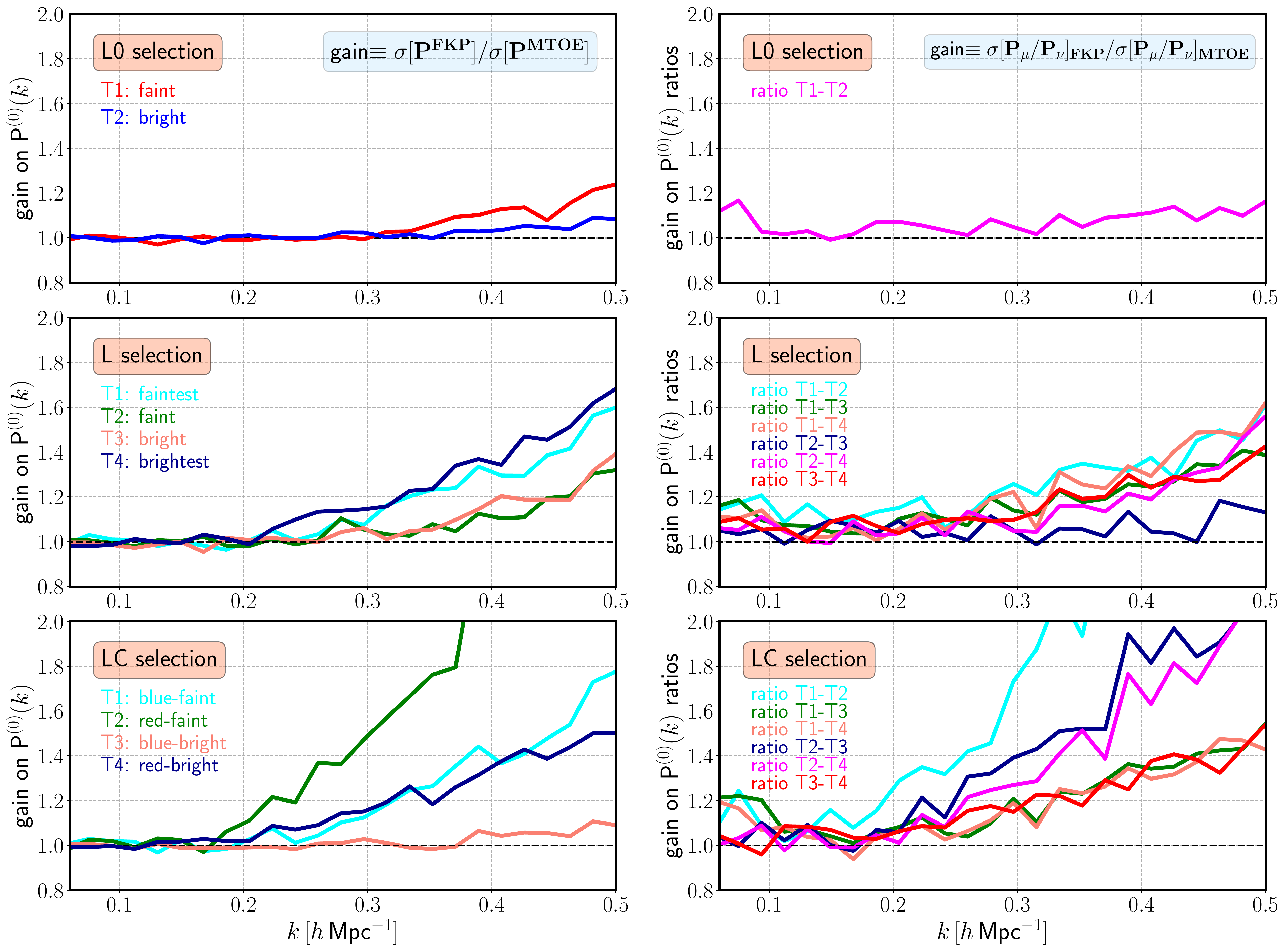}\hfill
\caption{The gain in accuracy provided by MTOE as compared to the standard FKP weighting for different tracer selections in the redshift range $0.55<z<0.7$. From top to bottom, increasingly more complicated selections are shown: L0 selection (2 tracers selected in luminosity), L-selection (4 tracers selected in luminosity) and LC-selection (4 tracers selected in colour and luminosity). In each row, the left-hand plot displays the gain for each individual $P^{(0)}(k)$ as a function of k, whereas the right-hand plot shows this effect on the ratios of individual spectra. Here, the gain is defined as the ratio between the corresponding standard deviations (taken from the diagonal of the sample covariance matrix).}
\label{fig:gains}
\end{center}
\end{figure*}

Importantly, the mean values of $P^{(0)}$(k) measured using the two estimates are almost identical within the entire range of scales considered, which confirms that MTOE is statistically unbiased with respect to the standard approach (see \citealt{Abramo2016}), even when the method is applied to a realistic mock catalogue. Note that differences are only barely noticeable in the highest redshift slice, and on very small scales (where uncertainties are very large). In addition, results from the two VIPERS fields are totally consistent with each other, with only small discrepancies that are also sub-dominant with respect to the estimated errors. Note that the volume covered by the W4 field is significantly smaller than that of the W1 field, which explains the larger errors associated with the W4 measurements.

MTOE provides a significantly more precise measurement of the monopoles $P^{(0)}$(k), especially on smaller scales. This effect can be glimpsed from Figure~\ref{fig:Pk}, but is more clearly illustrated by Figure~\ref{fig:corrmatrix_total}, which displays the complete correlation matrix for the 4 tracers considered as part of the LC selection in the first redshift slice ($0.6<z<0.75$), in the W1 field. The upper-diagonal terms correspond to the FKP correlation matrix, whereas the lower-diagonal ones show the MTOE measurement. Visually, it becomes obvious that the correlation between Fourier modes (the off-diagonal terms) is significantly higher in the FKP case compared with MTOE. The diagonal of the matrix, which allows us to compare the auto-correlations for each tracer as computed from MTOE and FKP, reveals that the MTOE measurement is also significantly more precise on scales $k\gtrsim0.25$. Although the strength of the effect depends on the particular tracer, redshift slice and field, the same qualitative behavior is found for all cases considered.

The fact that FKP is noisier than the MTOE approach on small scales is due to the influence of shot noise. The FKP estimator for the power spectrum of any given tracer only takes into account those tracers. The MTOE method, on the other hand, not only employs the auto-correlation of that tracer with itself, but also takes into account the cross-correlations of that tracer with all the other tracers in the estimation of the spectrum. Therefore, whereas the shot noise of the FKP estimator is the number density of that tracer, the {\it{effective shot noise}} of MTOE is the total number of tracers of the survey. Since shot noise affects more the small scales (where the amplitude of the spectrum is lower), the upshot is that MTOE is less noisy on small scales, effectively lowering the threshold imposed by shot noise. We have found qualitatively similar behaviors for the other redshift slices, and also when the simple L and L0 selections are imposed. We proceed now to quantify the magnitude of the effect.

In Figure~\ref{fig:gains}, we show the ratios of the relative standard deviations obtained from FKP and MTOE, as a function of $k$, for the first redshift slice in the W1 field. This quantity reflects the gain in accuracy, and is computed for both the individual power spectra (left-hand column) and the ratios of individual spectra (right-hand column). Each row in this figure represents a different tracer selection, in increasing order of complexity from top to bottom, i.e., L0, L, and LC, respectively. Several conclusions can be extracted from this figure, which are consistent with results from \cite{Abramo2016}. First, increasing the number of tracers (at least within some reasonable limits) tend to improve the performance of MTOE, as compared to FKP. Second, the advantages are observed both in terms of the individual power spectra, and on the ratios of individual spectra. Third, considering additional ways of discriminating between different populations is beneficial. This can be seen in the last row of Figure~\ref{fig:gains}, where colour is used as a second discriminator. Note that it is not the number of parameters what ultimately matters, but the combined discriminating capabilities of the parameter set. The average error reduction at $k\gtrsim0.3$ is around 40 $\%$ for the spectra and ratios in the W1 field, and 40-50 $\%$ in the W4 field. 

Figure~\ref{fig:gains} also reflects the most salient advantage of using multi-tracer techniques: the cancellation of cosmic variance on large scales. This effect can be seen most clearly in the ratios of spectra -- which MTOE has been shown to measure with higher accuracy \citep{Abramo2016}. The right-hand panels of Figure~\ref{fig:gains} show an average increase in the precision of this measurement of around $10\%$ at $k\lesssim0.1$ when MTOE is employed. The improvement is a little higher ($\sim15\%$) for the smaller W4 field.

The improvement that MTOE provides depends strongly on the number density of the particular tracer considered: the scarcer the tracer, the larger the improvement on individual spectra, and the smaller the scale at which the effect becomes significant. To demonstrate this, we draw attention to the left-bottom panel of Figure~\ref{fig:gains}, which displays the gain on individual spectra for the LC scheme. The number densities of the tracers are, respectively, 31 (T1), 5.6 (T2), 36.8 (T3), and 9.8 (T4) $10^{-4}$ [Mpc/h]$^{-3}$, as listed in Table~\ref{tab:schemes}. The scale at which the improvement reaches 20$\%$ is, respectively, $k\simeq$ 0.32 (T1), $\simeq$ 0.23 (T2), $>$ 0.5 (T3), $\simeq$ 0.32 (T4). Qualitatively, this general trend applies for any given selection and redshift slice - even though the particular gain will also depend on other factors, including the distribution of biases for a particular population.

In summary, the large increase in the signal-to-noise provided by MTOE on small scales for individual power spectra (and for the ratios of spectra) is a reflection of the method being better equipped to deal with shot noise, since MTOE combines information from all tracers. This feature is especially beneficial for sparse samples and on small scales, where the lack of signal becomes more severe. On the other hand, the improvement in the measurement of the ratios of spectra on large scales is due to the cancellation of cosmic variance, a feature that the MTOE is naturally designed to enhance.

\subsection{Derived quantities: Monte Carlo Markov Chain}
\label{sec:results_mcmc}

\begin{figure}
\begin{center}
\includegraphics[width=1\linewidth]{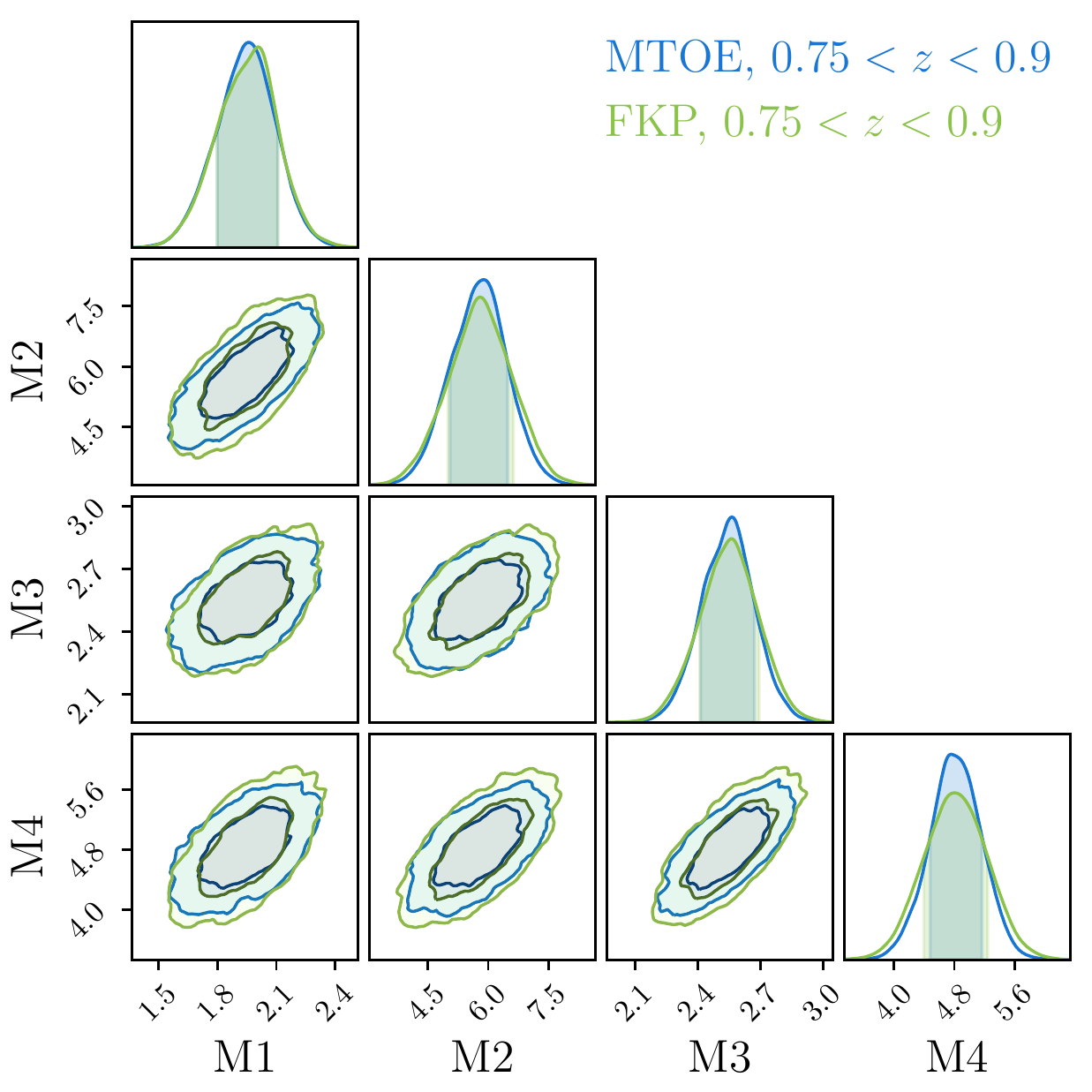}
\caption{Posterior probability distributions for the amplitudes of the monopoles (see Equation~\ref{eq:kaiser_monopole}) estimated using both MTOE and FKP for the 4 tracers considered as part of the LC selection in the redshift range $0.75<z<0.9$ (W1 field). The MCMC is performed within the wavenumber range $0.1 < k[h Mpc^{-1}] < 0.3$.}
\label{fig:monos}
\end{center}
\end{figure}

\begin{figure}
\begin{center}
\includegraphics[width=1.0\linewidth]{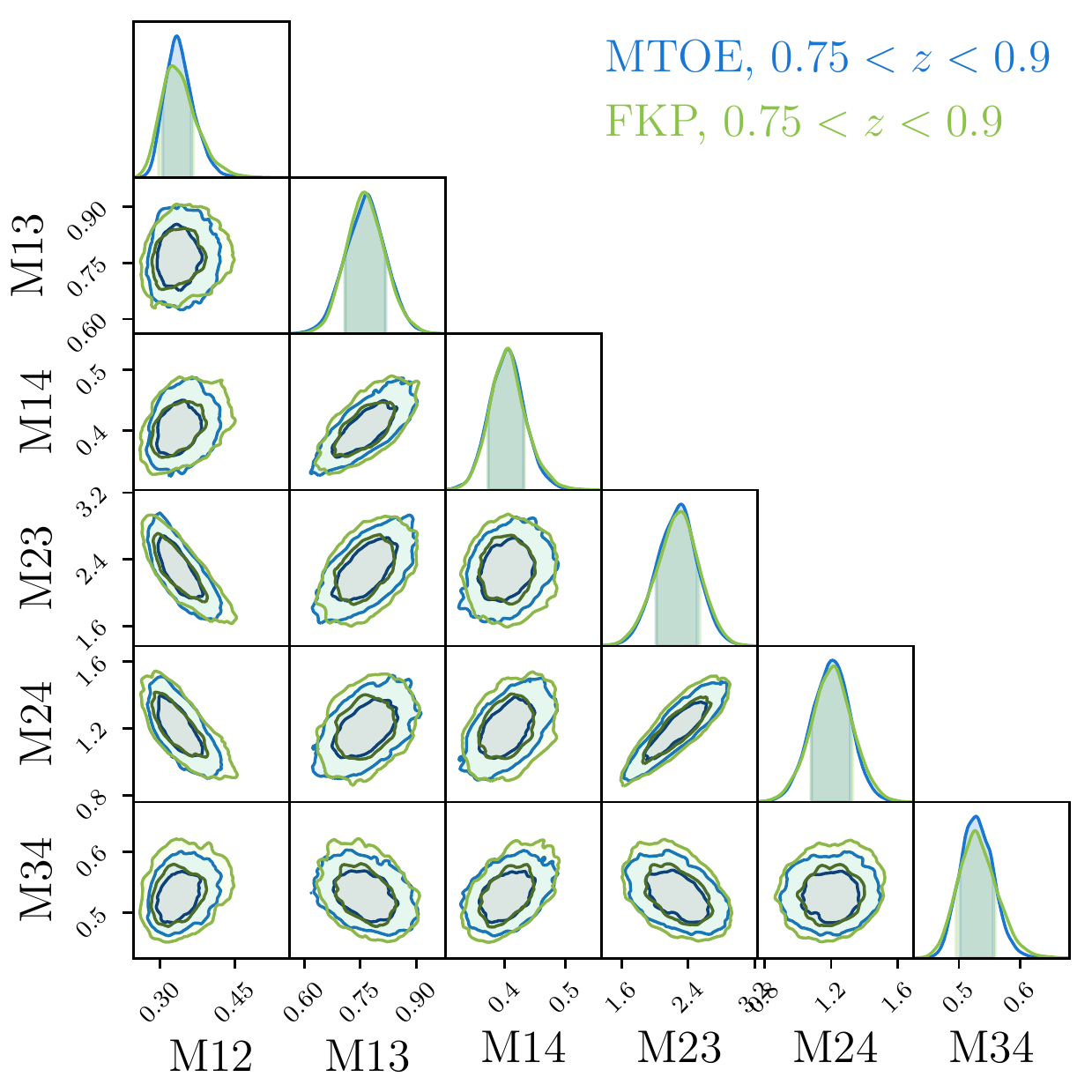}
\caption{Posterior probability distributions for the ratios of the monopoles for the 4 tracers considered as part of the LC selection in the redshift range $0.75<z<0.9$ (W1 field), estimated 
using both MTOE and FKP. The MCMC is performed within the wavenumber range $0.1 < k[h Mpc^{-1}] < 0.3$.}
\label{fig:monos_ratios}
\end{center}
\end{figure}

\begin{figure}
\begin{center}
\includegraphics[width=1\linewidth]{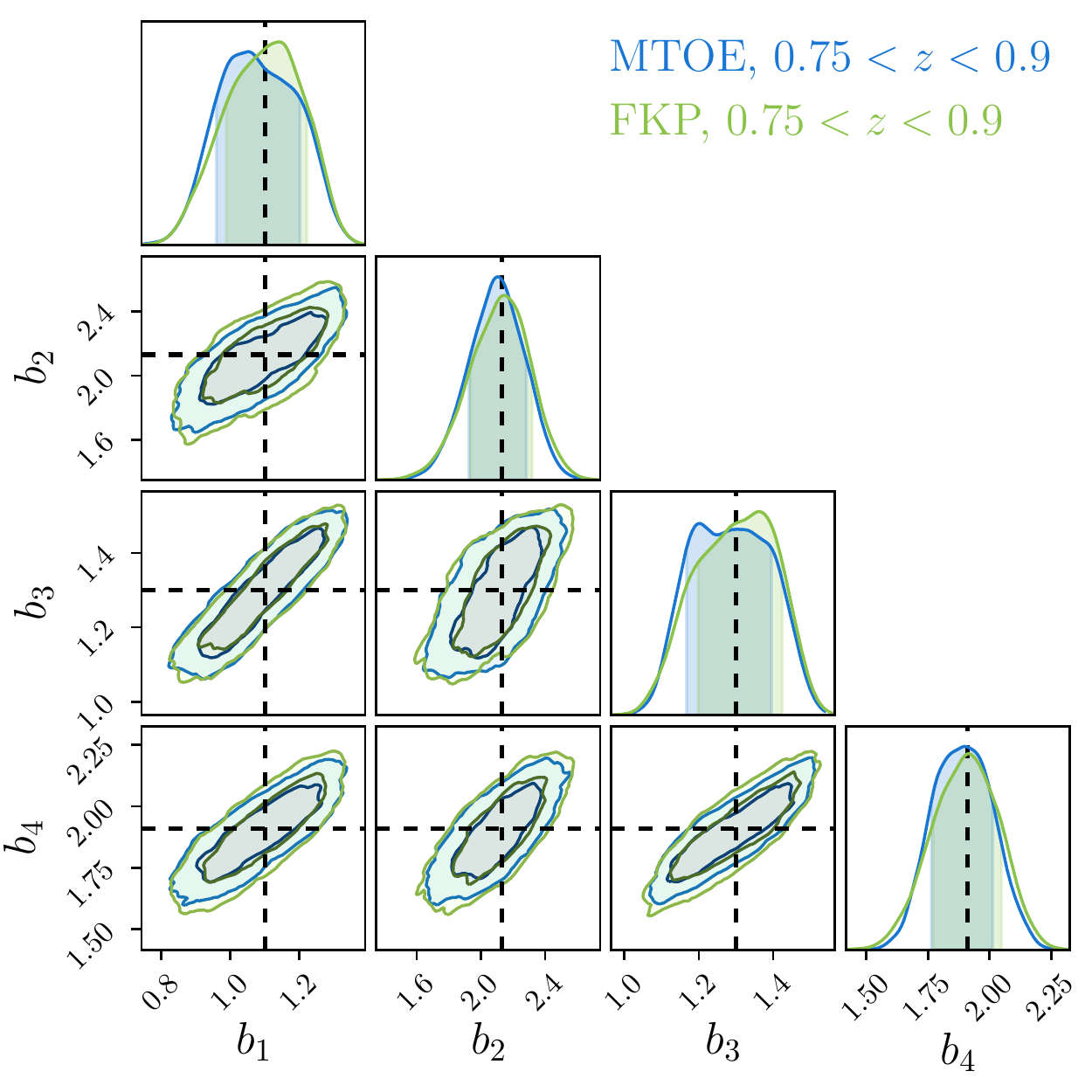}
\caption{Posterior probability distributions for the biases of the 4 tracers considered as part of the LC selection in the redshift range $0.75<z<0.9$ (W1 field), estimated using both FKP and MTOE. The MCMC is performed within the wavenumber range $0.1 < k[h Mpc^{-1}] < 0.3$.}
\label{fig:fb}
\end{center}
\end{figure}

The multipoles of the redshift-space power spectra constitute the basic measurements from which constraints on the galaxy biases and cosmological parameters can be obtained. Here we employ an MCMC algorithm to fit the normalized monopole $P^{(0)}$(k), and derive posterior probability distributions for the amplitudes of the monopoles, the ratios of the amplitudes of the monopoles, and the biases of tracers, at fixed cosmology. Since the main motivation for this analysis is to evaluate the performance of MTOE in a variety of configurations, we opt to present results from the W1 and W4 fields separately.

The monopole of the redshift-space power spectrum for a given tracer $\alpha$ is given by:

\be
\label{eq:kaiser}
P^{(0)}_\alpha(k) = M_\alpha P_m(k) \, ,
\\ \nonumber
\ee
where $M_\alpha$ is the first basic quantity that we fit from the power spectra of Figure~\ref{fig:Pk}, along with the ratios  $M_{\alpha\beta} = M_\alpha/M_\beta$. From the amplitudes $M_\alpha$, in linear order and in the flat-sky approximation (or, equivalently, plane parallel approximation, see, e.g., \citealt{Hamilton1998,Bertacca2012}), the bias of each tracer (b$_\alpha$) can be obtained assuming the following model \citep{Kaiser1987}:
\be
\label{eq:kaiser_monopole}
M_\alpha = b_\alpha^2+\frac{2}{3} f b_\alpha+ \frac{2}{5} f^2 \, ,
\ee
with $f$ being the matter growth function.

In this section, we show how the improvement on $P^{(0)}$(k) translates into $M_\alpha$, $M_{\alpha\beta}$, and $b_\alpha$, for all redshift slices and fields. A critical aspect in this computation is the integration range. As seen in Figure~\ref{fig:gains}, MTOE is especially advantageous on small scales. However, these are also the k-ranges where uncertainties, mode-couplings and non-linearities are larger, independently of the estimation method. Hence, we start by choosing a conservative range $0.1 < k[h Mpc^{-1}] < 0.3$.

\begin{figure}
\begin{center}
\includegraphics[width=1\linewidth]{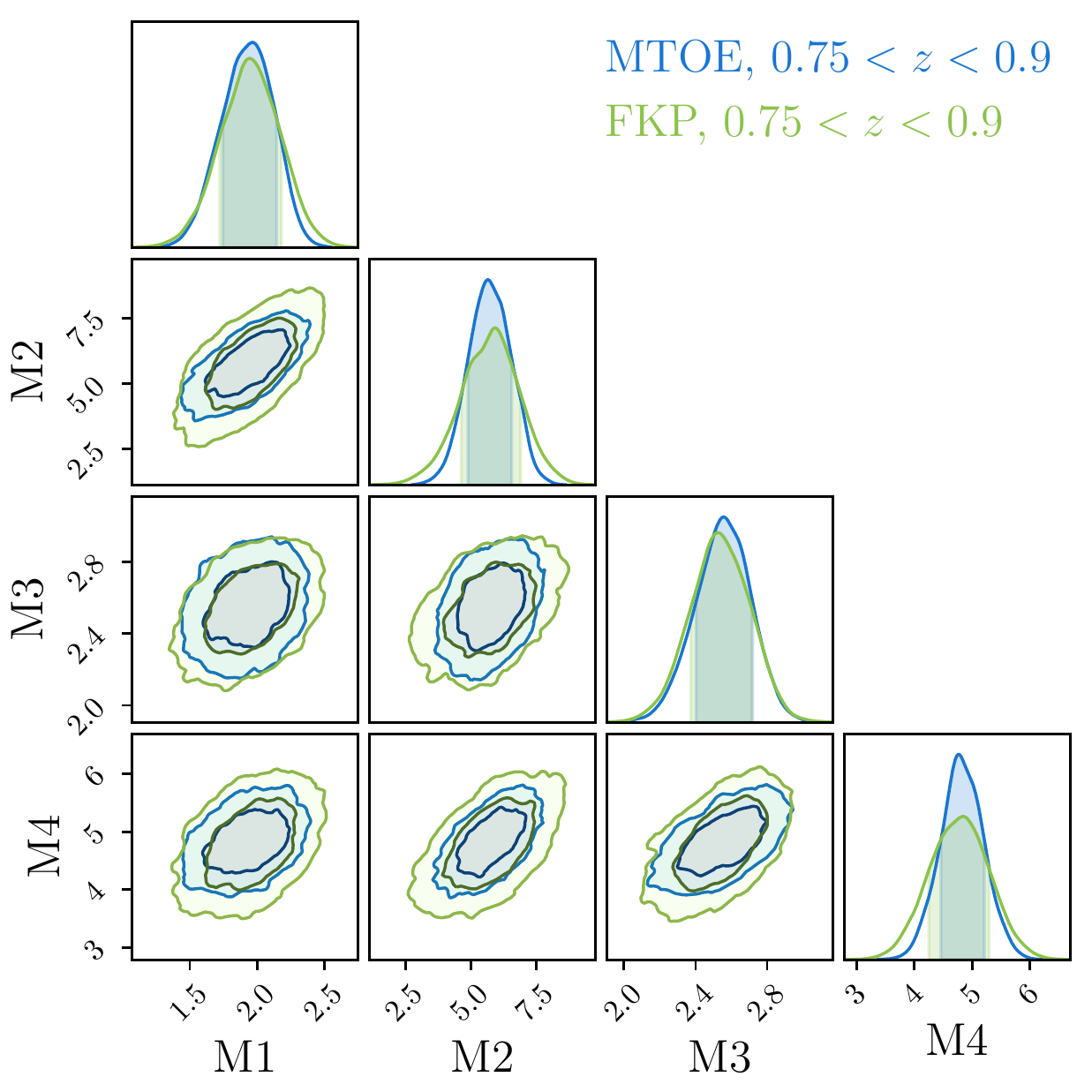}
\caption{The same as Figure~\ref{fig:monos} but assuming an integration range $0.3 < k[h Mpc^{-1}] < 0.5$.}
\label{fig:monos2}
\end{center}
\end{figure}

Figure~\ref{fig:monos} shows the posterior probability distributions for the amplitudes of the monopoles of the LC-selected tracers in the redshift range $0.75 < z < 0.9$ (second redshift slice). In this case, an additional (constant) shot-noise term was used as a nuisance parameter, but the cosmology was fixed. Results for the monopole ratios and for the biases are displayed in Figures~\ref{fig:monos_ratios} and~\ref{fig:fb}, respectively. Although the extent of the effect depends on the particular tracer, the improvement in the accuracy of the measurements, in terms of narrower posterior probability distributions, is observed, on average, for all parameters. 

\begin{figure*}
\begin{center}
\includegraphics[width=0.95\linewidth]{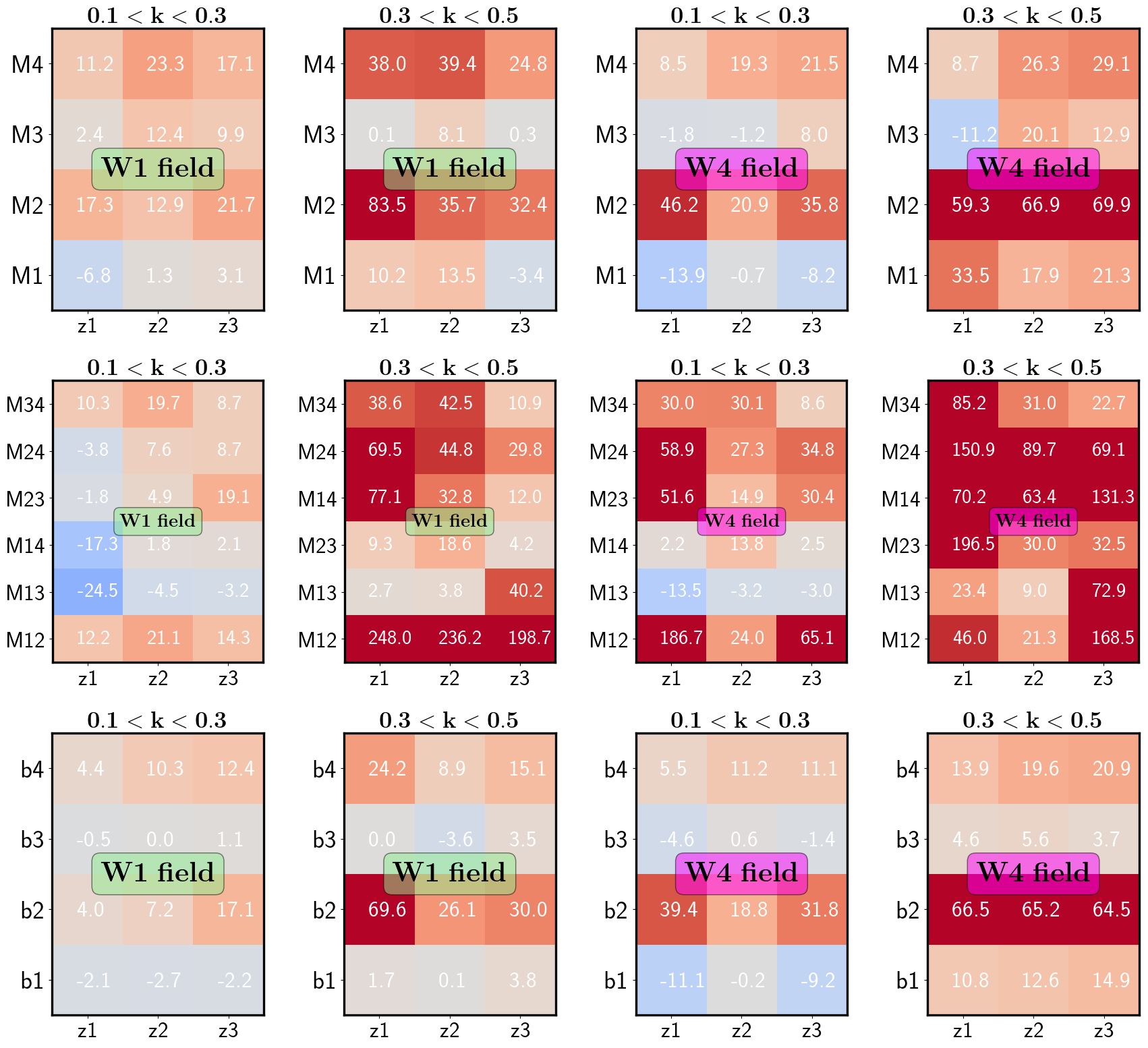}
\caption{The improvement provided by MTOE with respect to FKP for the amplitude of the monopoles, monopole ratios, and biases of the 4 tracers of the fiducial LC selection over the three redshift slices considered, in the W1 and W4 fields. The colour code corresponds to the percentage difference in the confidence intervals of the estimated parameters, i.e., $100\times (\sigma_{FKP}-\sigma_{MTOE})/\sigma_{MTOE}$. Redder colours indicate an increase in the accuracy, whereas bluer colours represent a decrease.     
The actual values of the percentage difference are provided inside each cell.}
\label{fig:gain_matrix}
\end{center}
\end{figure*}

The question that arises is how much we can further gain by pushing the measurements to higher values of $k$ -- a question that depends on the scientific purpose of the analysis. The answer to this question is illustrated in Figure~\ref{fig:monos2}, which displays the posterior probability distributions for the monopoles of the same tracers of Figure~\ref{fig:monos}, but now measured at the wavenumber range $0.3 < k[h Mpc^{-1}] < 0.5$. The difference between MTOE and FKP becomes apparent here, as expected from the improvements in the accuracy of the $P(k)$ measurement shown in Figure~\ref{fig:gains}.

In order to quantify the improvements, we define the ``gain'' provided by MTOE as the percentage difference between the confidence intervals derived from both methods, i.e., $100\times (\sigma_{FKP}-\sigma_{MTOE})/\sigma_{MTOE}$. In Figure~\ref{fig:gain_matrix}, we show these gains in a colour-coded diagram for all tracers and redshift slices, and for the two different integration ranges: $0.1 < k[h Mpc^{-1}] < 0.3$ and $0.3 < k[h Mpc^{-1}] < 0.5$. Redder colours indicate a positive gain (improvement), whereas bluer colours show a decrease in the accuracy of the measurement. The first thing to notice is that MTOE provides a more accurate measurement than FKP in the great majority of cases (80 $\%$). Also, the gains are notably larger for the much smaller W4 field. This is again a consequence of MTOE being more equipped to deal with low-number statistics and noisier samples.    

In the wavenumber range $0.1 < k[h Mpc^{-1}] < 0.3$, the gain is, on average, $\sim$11$\%$ for the amplitudes of the monopoles, $\sim$18$\%$ for the monopole ratios, and $\sim$6$\%$ for the biases (combining both fields). Individual cases, however, can reach much larger improvements. If we opt for a more aggressive take on small scales, the difference is boosted significantly. The average gain increases to $\sim$27$\%$ for the amplitudes, $\sim$70$\%$ for the monopole ratios, and $\sim$20$\%$ for the biases. In the Appendix, we list the mean values, confidence intervals and gains for all cases considered.

Since MTOE is mostly advantageous on small scales, for 
tracers with low number density, we must ensure that the method is robust for this configuration. As mentioned before, we have checked that the improvement shown in Figures~\ref{fig:gains}, ~\ref{fig:monos}, ~\ref{fig:monos_ratios}, ~\ref{fig:fb}, and ~\ref{fig:monos2} persists (qualitatively) when cells of 8 Mpc/h are employed (instead of 4 Mpc/h). In addition, we also computed the power spectra using a selection function smoothed with a Gaussian kernel, to check whether small-scale variations in the selection function could change the power -- and we verified that this had no effect whatsoever on our measurements.

\section{Discussion and conclusions}
\label{sec:discussion}

When multiple biased tracers occupy the same survey volume, the signal and the noise for all possible auto- and cross-correlations should be taken into account. The standard weighting scheme of FKP \citep*{FKP} was designed to be used in the context of the Fourier analysis of surveys based on a single population of LSS tracers. The PVP scheme \citep*{PVP}, on the other hand, provides optimal weights that lead to a minimum-variance estimator of the matter power spectrum $P_m(\mathbf{k})$ in situations where several different biased tracers are considered -- see also \citet{Cai2011}. The {\it{Multi-Tracer Optimal Estimator}} or MTOE \citep{Abramo2016}, which we describe and apply in this paper, is optimal both to estimate the matter power spectrum, as well as the redshift-space auto-power spectra of each individual tracer.

\begin{figure}
\includegraphics[width=1\linewidth]{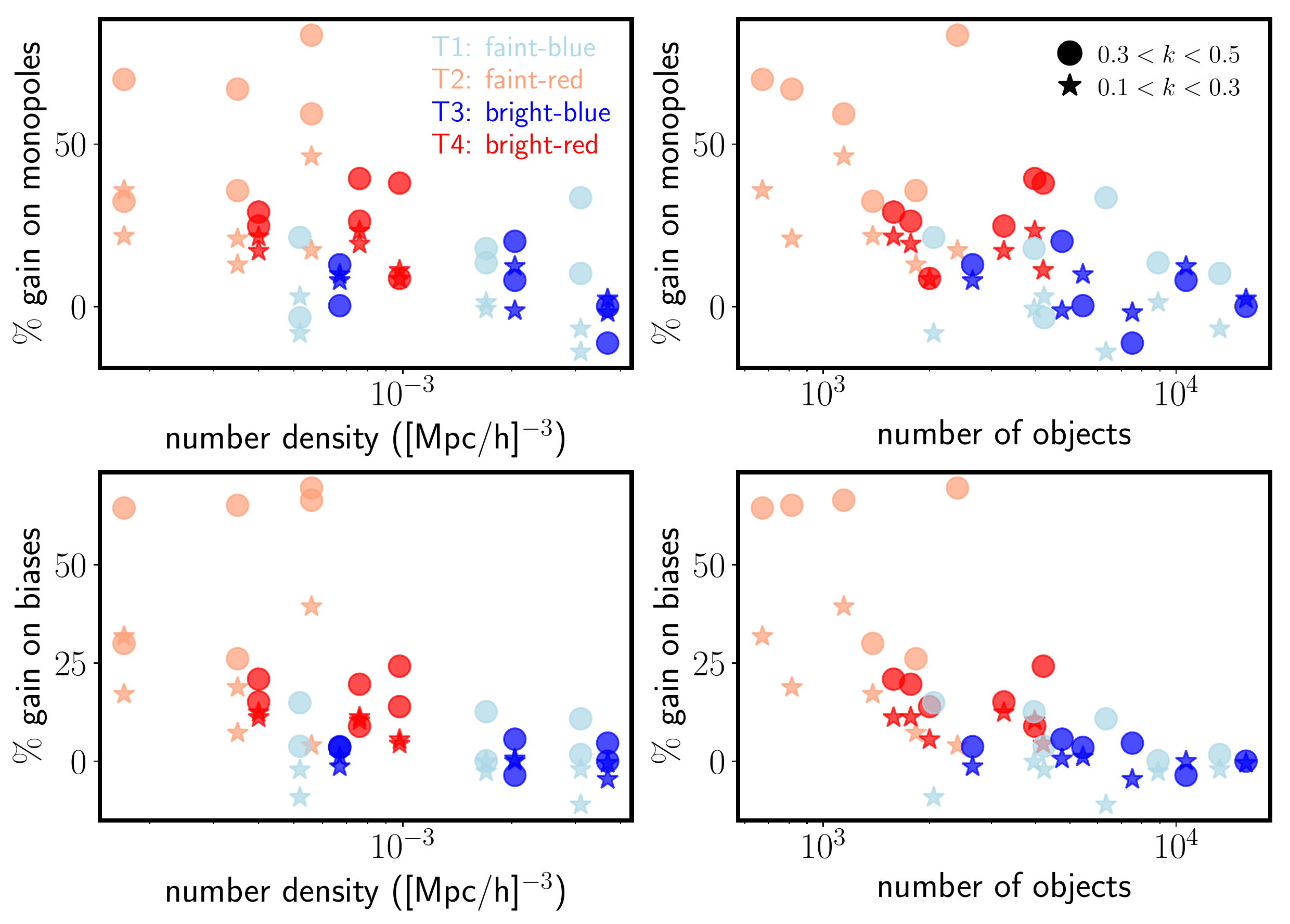}\hfill
\caption{The percentage gain in precision provided by MTOE with respect to FKP on the amplitude of the monopoles and biases as a function of the number density and total number of objects of the particular tracer sub-population. Here, tracers in different redshift slices and fields are shown in order to illustrate the dependence of the improvement on the abundance and type of tracer.}
\label{fig:summary}
\end{figure} 

The purpose of this work is threefold. First, to provide a compact, simplified and self-contained description of MTOE that can be easily implemented. Second, to evaluate the performance of the method on simulated but realistic galaxy data, and quantify the improvement with respect to the standard FKP approach. 
Third, to lay the foundations for the implementation of the method on the VIPERS data set, in order to improve constraints not only on the power spectra and galaxy biases, but also on the cosmological parameters. 

On the theoretical side, the description of the method that we provide incorporates an analytic derivation of the window functions of the auto-power spectra -- a result that was not known previously. 
We show that the MTOE window functions conserve the total power of the measurement: they are normalized to unity in terms of the modes, and the mean mixing between the different tracers vanishes when averaged over all the modes.

We have shown that MTOE provides significant advantages when applied to a data set split in multiple galaxy populations. It is unbiased with respect to the standard FKP approach, and improves upon its performance. In the case of VIPERS, this gain is more pronounced on small scales. We report an increase in the signal-to-noise of the monopoles of the auto-power spectra of $\sim$ 40$\%$ at $k[h Mpc^{-1}] \gtrsim 0.3$, with improvements increasing towards smaller scales -- the specific values depending on the particular tracer and on the characteristics of the survey. The improvements on small scales are also large for the ratios of power spectra. We have also implemented an MCMC procedure in order to explore the probability density space of derived fitting quantities such as the amplitudes of the monopoles, the monopole ratios, and the galaxy biases. If we can push the measurement to a wavenumber range $0.3 < k[h Mpc^{-1}] < 0.5$, the gain on these quantities reaches, on average, 30, 70, and 20 $\%$, respectively.

The particular extent of the improvement for a given tracer sub-sample is determined, to a large degree, by the survey volume, the number density, and the bias of the tracer. We illustrate this with a summary plot in Figure~\ref{fig:summary}. Here, tracers from different selections and redshifts are shown together, and the corresponding gains plotted as a function of number density and number of objects. For abundant tracers, the advantage of using MTOE is very mild. For scarcer tracer, the gain is very significant, depending strongly on the type of tracer: redder tracers, which are more strongly bias, benefit more from the MTOE weighting.

The improvement provided by MTOE is not restricted to small scales. As expected from the very foundation of the method, which is based on combining different measurements in order minimize cosmic variance, we find an average gain of $\sim$ 10$\%$ on scales $k[h Mpc^{-1}] \lesssim 0.1$ for the ratios of spectra. Importantly, the application of the method to larger cosmological volumes in the future will allow us to test whether we can achieve larger improvements on even larger scales.

The second paper of this series will focus on the application of the method to the VIPERS data set, as well as the inclusion of all cross-spectra of the tracers. In particular, it is still to be determined how much of the reported gains can be recovered when other sources of uncertainties, associated with a real data set, are present. The most relevant issue that we can anticipate is the fidelity of the mocks with respect to the data, in the context of a multi-tracer analysis. It is important to stress that MTOE can be applied on data only without mocks. However, mocks are needed to estimate errors. 
Mocks (including the VIPERS mocks) are typically built to reproduce the clustering of a combined data set, or, in some cases, the dependence of clustering on stellar mass or luminosity. As we have shown, the ideal situation for a multi-tracer approach involves having multiple ways of selecting galaxy populations (e..g., luminosity and colour). If mocks do not reproduce the bias and number density of these sub-populations, the derived errors can be inaccurate. Accurate ``multi-tracer mocks" are thus needed.

This paper shows that MTOE is also a powerful tool to explore the physics that takes place on small cosmological scales from the perspective of the power spectrum measurement. On scales $\lesssim 10$ Mpc/h, the effect of the 1-halo term \citep{Cooray2002}, which describes the clustering of galaxies inside haloes, starts to become dominant and can become degenerate with with shot noise. MTOE provides a route into the astrophysics of multiple galaxy populations, since it can be used to place tighter constraints on halo occupation distribution (HOD) models on small scales. 

From the perspective of the halo-galaxy connection, improving on our ability to measure galaxy biases is advantageous in the context of assembly bias studies (see, e.g., \citealt{Miyatake2016,Lin2016,MonteroDorta2017, Niemiec2018}). In \cite{SatoPolito2018} we show how applying the MTOE to multiple subsets of haloes increases the signal-to-noise of the secondary bias (i.e., ``halo assembly bias") measurement to a level that can only be achieved if the underlying dark-matter density field of the simulation is known. This convenient ``trick" is not attainable with real data, therefore the usefulness of MTOE. 

The unprecedented wealth of cosmological survey data that will soon be available to the community motivates us to continue developing MTOE. Experiments such as J-PAS, DESI, Euclid, or the LSST will map huge volumes of the high-redshift universe with broad galaxy selections, seeking accurate cosmological measurements over unexplored redshift ranges. MTOE and other multi-tracer approaches, applied to a single data set or a combination of multiple surveys, can help us push the limits of the clustering measurement to scales that were not accessible before.

\section*{Acknowledgments}

ADMD thanks FAPESP for financial support. LRA thanks both FAPESP and CNPq for financial support. BRG and LG acknowledge support of ASI, through contract n 2018-23-HH.0 ``Euclid".
SdlT acknowledges the support of the OCEVU Labex (ANR-11-LABX-0060) and the A*MIDEX project (ANR-11-IDEX-0001-02) funded by the ``Investissements d'Avenir" French government program managed by the ANR.

We also thank S. Vitenti, M. Penna-Lima and C. Doux for the use of their Numerical Cosmology package, {\em{NumCosmo}}\footnote{\url{https://numcosmo.github.io/}}.

\bibliography{./paper}

\begin{thebibliography}{58}
\expandafter\ifx\csname natexlab\endcsname\relax\def\natexlab#1{#1}\fi

\bibitem[{Abramo}(2012)]{Abramo2012}
{Abramo} L.~R., 2012, \mnras, 420, 2042

\bibitem[{Abramo} \& {Bertacca}(2017)]{Abramo2017}
{Abramo} L.~R., {Bertacca} D., 2017, \prd, 96, 12, 123535

\bibitem[{Abramo} \& {Leonard}(2013)]{Abramo2013}
{Abramo} L.~R., {Leonard} K.~E., 2013, \mnras, 432, 318

\bibitem[{Abramo} et~al.(2016){Abramo}, {Secco} \& {Loureiro}]{Abramo2016}
{Abramo} L.~R., {Secco} L.~F., {Loureiro} A., 2016, \mnras, 455, 3871

\bibitem[Alarcon et~al.(2018)Alarcon, Eriksen \& Gaztañaga]{Alarcon2016}
Alarcon A., Eriksen M., Gaztañaga E., 2018, Mon. Not. Roy. Astron. Soc., 473,
  2, 1444

\bibitem[{Benoist} et~al.(1996){Benoist}, {Maurogordato}, {da Costa}, {Cappi}
  \& {Schaeffer}]{Benoist1996}
{Benoist} C., {Maurogordato} S., {da Costa} L.~N., {Cappi} A., {Schaeffer} R.,
  1996, \apj, 472, 452

\bibitem[{Bertacca} et~al.(2012){Bertacca}, {Maartens}, {Raccanelli} \&
  {Clarkson}]{Bertacca2012}
{Bertacca} D., {Maartens} R., {Raccanelli} A., {Clarkson} C., 2012, \jcap,
  2012, 10, 025

\bibitem[{Bianchi} et~al.(2015){Bianchi}, {Gil-Mar{\'{\i}}n}, {Ruggeri} \&
  {Percival}]{Bianchi2015}
{Bianchi} D., {Gil-Mar{\'{\i}}n} H., {Ruggeri} R., {Percival} W.~J., 2015,
  \mnras, 453, L11

\bibitem[{Blake} et~al.(2013){Blake}, {Baldry}, {Bland-Hawthorn}
  et~al.]{Blake2013}
{Blake} C., {Baldry} I.~K., {Bland-Hawthorn} J., et~al., 2013, \mnras, 436,
  3089

\bibitem[{Bull} et~al.(2015){Bull}, {Ferreira}, {Patel} \& {Santos}]{Bull2015}
{Bull} P., {Ferreira} P.~G., {Patel} P., {Santos} M.~G., 2015, \apj, 803, 21

\bibitem[{Cai} \& {Bernstein}(2012)]{Cai2012}
{Cai} Y.-C., {Bernstein} G., 2012, \mnras, 422, 1045

\bibitem[{Cai} et~al.(2011){Cai}, {Bernstein} \& {Sheth}]{Cai2011}
{Cai} Y.-C., {Bernstein} G., {Sheth} R.~K., 2011, \mnras, 412, 995

\bibitem[{Coles} \& {Jones}(1991)]{Coles1991}
{Coles} P., {Jones} B., 1991, \mnras, 248, 1

\bibitem[{Cooray} \& {Sheth}(2002)]{Cooray2002}
{Cooray} A., {Sheth} R., 2002, \physrep, 372, 1

\bibitem[{Davis} \& {Huchra}(1982)]{Davis1982}
{Davis} M., {Huchra} J., 1982, \apj, 254, 437

\bibitem[{Dawson} et~al.(2013){Dawson}, {Schlegel}, {Ahn} et~al.]{Dawson2013}
{Dawson} K.~S., {Schlegel} D.~J., {Ahn} C.~P., et~al., 2013, \aj, 145, 10

\bibitem[{de la Torre} et~al.(2013){de la Torre}, {Guzzo}, {Peacock}
  et~al.]{DeLaTorre2013V}
{de la Torre} S., {Guzzo} L., {Peacock} J.~A., et~al., 2013, \aap, 557, A54

\bibitem[{de la Torre} et~al.(2017){de la Torre}, {Jullo}, {Giocoli}
  et~al.]{DeLaTorre2017}
{de la Torre} S., {Jullo} E., {Giocoli} C., et~al., 2017, \aap, 608, A44

\bibitem[{de la Torre} \& {Peacock}(2013)]{DeLaTorre2013}
{de la Torre} S., {Peacock} J.~A., 2013, \mnras, 435, 743

\bibitem[{Eisenstein} et~al.(2011){Eisenstein}, {Weinberg}, {Agol}
  et~al.]{Eisenstein2011}
{Eisenstein} D.~J., {Weinberg} D.~H., {Agol} E., et~al., 2011, \aj, 142, 72

\bibitem[{Feldman} et~al.(1994){Feldman}, {Kaiser} \& {Peacock}]{FKP}
{Feldman} H.~A., {Kaiser} N., {Peacock} J.~A., 1994, \apj, 426, 23

\bibitem[{Ferramacho} et~al.(2014){Ferramacho}, {Santos}, {Jarvis} \&
  {Camera}]{Ferramacho2014}
{Ferramacho} L.~D., {Santos} M.~G., {Jarvis} M.~J., {Camera} S., 2014, \mnras,
  442, 2511

\bibitem[Fonseca et~al.(2015)Fonseca, Camera, Santos \& Maartens]{Fonseca2015}
Fonseca J., Camera S., Santos M., Maartens R., 2015, Astrophys. J., 812, 2, L22

\bibitem[{Garilli} et~al.(2014){Garilli}, {Guzzo}, {Scodeggio}
  et~al.]{Garilli2014}
{Garilli} B., {Guzzo} L., {Scodeggio} M., et~al., 2014, \aap, 562, A23

\bibitem[{Gil-Mar{\'{\i}}n} et~al.(2010){Gil-Mar{\'{\i}}n}, {Wagner}, {Verde},
  {Jimenez} \& {Heavens}]{GilMarin2010}
{Gil-Mar{\'{\i}}n} H., {Wagner} C., {Verde} L., {Jimenez} R., {Heavens} A.~F.,
  2010, \mnras, 407, 772

\bibitem[{Granett} et~al.(2015){Granett}, {Branchini}, {Guzzo}
  et~al.]{Granett2015}
{Granett} B.~R., {Branchini} E., {Guzzo} L., et~al., 2015, \aap, 583, A61

\bibitem[{Guzzo} et~al.(2014){Guzzo}, {Scodeggio}, {Garilli} et~al.]{Guzzo2014}
{Guzzo} L., {Scodeggio} M., {Garilli} B., et~al., 2014, \aap, 566, A108

\bibitem[{Guzzo} et~al.(1997){Guzzo}, {Strauss}, {Fisher}, {Giovanelli} \&
  {Haynes}]{Guzzo1997}
{Guzzo} L., {Strauss} M.~A., {Fisher} K.~B., {Giovanelli} R., {Haynes} M.~P.,
  1997, \apj, 489, 37

\bibitem[{Hamaus} et~al.(2011){Hamaus}, {Seljak} \& {Desjacques}]{Hamaus2011}
{Hamaus} N., {Seljak} U., {Desjacques} V., 2011, \prd, 84, 8, 083509

\bibitem[{Hamaus} et~al.(2012){Hamaus}, {Seljak} \& {Desjacques}]{Hamaus2012}
{Hamaus} N., {Seljak} U., {Desjacques} V., 2012, \prd, 86, 10, 103513

\bibitem[{Hamilton}(1993)]{Hamilton1993}
{Hamilton} A.~J.~S., 1993, The Astrophysical Journal, 417, 19

\bibitem[{Hamilton}(1998)]{Hamilton1998}
{Hamilton} A.~J.~S., 1998, in { The Evolving Universe\/}, edited by
  D.~{Hamilton}, vol. 231 of { Astrophysics and Space Science Library\/},  185

\bibitem[{Jing}(2005)]{Jing2005}
{Jing} Y.~P., 2005, \apj, 620, 559

\bibitem[{Kaiser}(1987)]{Kaiser1987}
{Kaiser} N., 1987, \mnras, 227, 1

\bibitem[{Klypin} et~al.(2016){Klypin}, {Yepes}, {Gottl{\"o}ber}, {Prada} \&
  {He{\ss}}]{Klypin2016}
{Klypin} A., {Yepes} G., {Gottl{\"o}ber} S., {Prada} F., {He{\ss}} S., 2016,
  \mnras, 457, 4340

\bibitem[{Lewis} \& {Bridle}(2002)]{CAMB}
{Lewis} A., {Bridle} S., 2002, \prd, 66, 10, 103511

\bibitem[{Lin} et~al.(2016){Lin}, {Mandelbaum}, {Huang} et~al.]{Lin2016}
{Lin} Y.-T., {Mandelbaum} R., {Huang} Y.-H., et~al., 2016, \apj, 819, 119

\bibitem[{McDonald}(2008)]{McDonald2008}
{McDonald} P., 2008, \prd, 78, 12, 123519

\bibitem[{Miyatake} et~al.(2016){Miyatake}, {More}, {Takada}
  et~al.]{Miyatake2016}
{Miyatake} H., {More} S., {Takada} M., et~al., 2016, Physical Review Letters,
  116, 4, 041301

\bibitem[{Mohammad} et~al.(2018){Mohammad}, {Bianchi}, {Percival}
  et~al.]{Mohammad2018}
{Mohammad} F.~G., {Bianchi} D., {Percival} W.~J., et~al., 2018, \aap, 619, A17

\bibitem[{Montero-Dorta} et~al.(2017){Montero-Dorta}, {P{\'e}rez}, {Prada}
  et~al.]{MonteroDorta2017}
{Montero-Dorta} A.~D., {P{\'e}rez} E., {Prada} F., et~al., 2017, \apjl, 848, L2

\bibitem[{Navarro} et~al.(1997){Navarro}, {Frenk} \& {White}]{NFW}
{Navarro} J.~F., {Frenk} C.~S., {White} S.~D.~M., 1997, \apj, 490, 493

\bibitem[{Niemiec} et~al.(2018){Niemiec}, {Jullo}, {Montero-Dorta}
  et~al.]{Niemiec2018}
{Niemiec} A., {Jullo} E., {Montero-Dorta} A.~D., et~al., 2018, \mnras, 477, L1

\bibitem[{Percival} et~al.(2004){Percival}, {Verde} \& {Peacock}]{PVP}
{Percival} W.~J., {Verde} L., {Peacock} J.~A., 2004, \mnras, 347, 645

\bibitem[{Pezzotta} et~al.(2017){Pezzotta}, {de la Torre}, {Bel}
  et~al.]{Pezzotta2017}
{Pezzotta} A., {de la Torre} S., {Bel} J., et~al., 2017, \aap, 604, A33

\bibitem[{Planck Collaboration} et~al.(2014){Planck Collaboration}, {Ade},
  {Aghanim} et~al.]{planck2014}
{Planck Collaboration}, {Ade} P.~A.~R., {Aghanim} N., et~al., 2014, \aap, 571,
  A1

\bibitem[{Rota} et~al.(2017){Rota}, {Granett}, {Bel} et~al.]{Rota2017}
{Rota} S., {Granett} B.~R., {Bel} J., et~al., 2017, \aap, 601, A144

\bibitem[{Sato-Polito} et~al.(2019){Sato-Polito}, {Montero-Dorta}, {Abramo},
  {Prada} \& {Klypin}]{SatoPolito2018}
{Sato-Polito} G., {Montero-Dorta} A.~D., {Abramo} L.~R., {Prada} F., {Klypin}
  A., 2019, \mnras, 487, 2, 1570

\bibitem[{Scoccimarro}(2015)]{Scoccimarro2015}
{Scoccimarro} R., 2015, \prd, 92, 8, 083532

\bibitem[{Scodeggio} et~al.(2018){Scodeggio}, {Guzzo}, {Garilli}
  et~al.]{Scodeggio2018}
{Scodeggio} M., {Guzzo} L., {Garilli} B., et~al., 2018, \aap, 609, A84

\bibitem[{Seljak}(2009)]{Seljak2009}
{Seljak} U., 2009, Physical Review Letters, 102, 2, 021302

\bibitem[{Skibba} et~al.(2006){Skibba}, {Sheth}, {Connolly} \&
  {Scranton}]{Skibba2006}
{Skibba} R., {Sheth} R.~K., {Connolly} A.~J., {Scranton} R., 2006, \mnras, 369,
  68

\bibitem[{Skibba}(2009)]{Skibba2009}
{Skibba} R.~A., 2009, \mnras, 392, 1467

\bibitem[{Slosar}(2009)]{Slosar2009}
{Slosar} A., 2009, \jcap, 3, 004

\bibitem[{Smith} et~al.(2011){Smith}, {Desjacques} \& {Marian}]{Smith2011}
{Smith} R.~E., {Desjacques} V., {Marian} L., 2011, \prd, 83, 4, 043526

\bibitem[{Smith} et~al.(2003){Smith}, {Peacock}, {Jenkins} et~al.]{HaloFit}
{Smith} R.~E., {Peacock} J.~A., {Jenkins} A., et~al., 2003, \mnras, 341, 1311

\bibitem[{Tegmark} et~al.(1998){Tegmark}, {Hamilton}, {Strauss}, {Vogeley} \&
  {Szalay}]{TegmarkHamilton1998}
{Tegmark} M., {Hamilton} A.~J.~S., {Strauss} M.~A., {Vogeley} M.~S., {Szalay}
  A.~S., 1998, \apj, 499, 555

\bibitem[{Witzemann} et~al.(2019){Witzemann}, {Alonso}, {Fonseca} \&
  {Santos}]{Witzemann2018}
{Witzemann} A., {Alonso} D., {Fonseca} J., {Santos} M.~G., 2019, \mnras, 485,
  4, 5519

\end{thebibliography}

\appendix
\section{Best-fitting values}

We provide here best-fitting values for the amplitudes of the monopoles, monopole ratios and biases of the LC tracers estimated using both the FKP and MTOE methods, in the three redshift slices considered. Table~\ref{tab:gains1} and Table~\ref{tab:gains2} displays results for integration ranges of $0.1 < k[Mpc/h] < 0.3$ and $0.3 < k[Mpc/h] < 0.5$, respectively, in the W1 field. Results for the W4 field are listed in Table~\ref{tab:gains3} and Table~\ref{tab:gains4}. 


\begin{table*}      
\caption{The best-fitting FKP and MTOE estimates of the amplitudes of the monopoles, monopole ratios and biases of the LC tracers along with the percentage reduction of the uncertainty provided by MTOE ($100\times \sigma_{FKP}-\sigma_{MTOE}/\sigma_{MTOE}$), for an integration range $0.1 < k[Mpc/h] < 0.3$, in the W1 field.}    
\begin{center}
\label{tab:gains1}   
\begin{tabular}{c | ccc | ccc | ccc}   
\hline &     \multicolumn{3}{ |c| }{$0.6<z<0.75$}& \multicolumn{3}{ |c| }{$0.75<z<0.9$}& \multicolumn{3}{ |c| }{$0.9<z<1.1$} \\
\hline  Param. &   MTOE& FKP& Gain($\%$)&  MTOE& FKP& Gain($\%$)& MTOE& FKP& Gain($\%$)\\
\hline 
M1& 1.671$\pm$0.145& 1.629$\pm$0.136& {\bf{-6.8}}&   1.948$\pm$0.151& 2.010$\pm$0.153& {\bf{+1.2}}&   2.367$\pm$0.251& 2.345$\pm$0.259& {\bf{+3.1}}  \\
M2& 4.838$\pm$0.569& 4.947$\pm$0.667& {\bf{+17.3}}&  5.920$\pm$0.708& 5.754$\pm$0.799& {\bf{+12.9}}&      5.739$\pm$0.564& 5.603$\pm$0.686& {\bf{+21.7}}  \\
M3& 1.990$\pm$0.098& 1.961$\pm$0.100& {\bf{+2.4}}&   2.572$\pm$0.126& 2.564$\pm$0.142& {\bf{+12.4}}&      3.265$\pm$0.192& 3.273$\pm$0.210& {\bf{+9.9}} \\
M4& 4.404$\pm$0.330& 4.293$\pm$0.367& {\bf{+11.2}}&  4.722$\pm$0.342& 4.854$\pm$0.421& {\bf{+23.3}}&      5.352$\pm$0.351& 5.386$\pm$0.411& {\bf{+17.1}} \\
\hline 
M$_{1,2}$& 0.332$\pm$0.031& 0.330$\pm$0.035& {\bf{+12.2}}&  0.332$\pm$0.028& 0.322$\pm$0.034& {\bf{+21.1}}&   0.412$\pm$0.040& 0.401$\pm$0.046& {\bf{+14.3}}  \\
M$_{1,3}$& 0.851$\pm$0.078& 0.822$\pm$0.059& {\bf{-24.5}}&  0.767$\pm$0.054& 0.759$\pm$0.051& {\bf{-4.5}}&   0.722$\pm$0.070& 0.716$\pm$0.068& {\bf{-3.2}} \\
M$_{1,4}$& 0.366$\pm$0.044& 0.367$\pm$0.036& {\bf{-17.3}}&  0.406$\pm$0.028& 0.405$\pm$0.029& {\bf{+1.8}}&   0.431$\pm$0.048& 0.446$\pm$0.050& {\bf{+2.1}} \\
M$_{2,3}$& 2.451$\pm$0.292& 2.366$\pm$0.287& {\bf{-1.8}}&   2.314$\pm$0.245& 2.339$\pm$0.257& {\bf{+4.9}}&   1.764$\pm$0.158& 1.749$\pm$0.189& {\bf{+19.1}} \\
M$_{2,4}$& 1.105$\pm$0.139& 1.128$\pm$0.134& {\bf{-3.8}}&   1.192$\pm$0.115& 1.220$\pm$0.124& {\bf{+7.6}}&   1.061$\pm$0.102& 1.049$\pm$0.111& {\bf{+8.7}} \\
M$_{3,4}$& 0.443$\pm$0.024& 0.453$\pm$0.027& {\bf{+10.3}}&  0.528$\pm$0.026& 0.526$\pm$0.032& {\bf{+19.7}}&   0.599$\pm$0.037& 0.603$\pm$0.040& {\bf{+8.7}}\\
\hline 
b$_{1}$& 0.952$\pm$0.116& 0.941$\pm$0.114& {\bf{-2.1}}&  1.059$\pm$0.119& 1.155$\pm$0.116& {\bf{-2.7}}&  1.197$\pm$0.137& 1.217$\pm$0.134& {\bf{-2.2}} \\
b$_{2}$& 1.911$\pm$0.164& 1.877$\pm$0.171& {\bf{+4.0}}&  2.111$\pm$0.177& 2.147$\pm$0.190& {\bf{+7.2}}&  2.048$\pm$0.153& 2.089$\pm$0.180& {\bf{+17.1}}  \\
b$_{3}$& 1.126$\pm$0.105& 1.134$\pm$0.104& {\bf{-0.5}}&  1.202$\pm$0.0& 1.364$\pm$0.112& {\bf{--}}&      1.408$\pm$0.118& 1.524$\pm$0.119& {\bf{+1.1}} \\
b$_{4}$& 1.782$\pm$0.117& 1.790$\pm$0.122& {\bf{+4.4}}&  1.891$\pm$0.123& 1.912$\pm$0.136& {\bf{+10.2}}&  2.008$\pm$0.123& 1.992$\pm$0.138& {\bf{+12.4}} \\

\hline  

\end{tabular}
\end{center}
\end{table*}

\begin{table*}      
\caption{Same as Table~\ref{tab:gains1} for an integration range $0.3 < k[Mpc/h] < 0.5$ (W1 field).}    
\begin{center}
\label{tab:gains2}   
\begin{tabular}{c | ccc | ccc | ccc}   
\hline &     \multicolumn{3}{ |c| }{$0.6<z<0.75$}& \multicolumn{3}{ |c| }{$0.75<z<0.9$}& \multicolumn{3}{ |c| }{$0.9<z<1.1$} \\
\hline  Param. &   MTOE& FKP& Gain($\%$)&  MTOE& FKP& Gain($\%$)& MTOE& FKP& Gain($\%$)\\
\hline 
M1& 1.675$\pm$0.154& 1.600$\pm$0.170& {\bf{+10.2}}& 1.981$\pm$0.198& 1.932$\pm$0.225& {\bf{+13.5}}&   2.297$\pm$0.387& 2.269$\pm$0.374& {\bf{-3.4}} \\
M2& 4.828$\pm$0.561& 4.633$\pm$1.029& {\bf{+83.5}}& 5.639$\pm$0.829& 5.913$\pm$1.125& {\bf{+35.7}}&    5.765$\pm$0.944& 5.726$\pm$1.250& {\bf{+32.4}}   \\
M3& 1.997$\pm$0.098& 1.974$\pm$0.098& {\bf{+0.1}}&   2.558$\pm$0.156& 2.515$\pm$0.169& {\bf{+8.1}}&    3.274$\pm$0.307& 3.239$\pm$0.307& {\bf{+0.3}} \\
M4& 4.480$\pm$0.318& 4.387$\pm$0.439& {\bf{+38.0}}&   4.746$\pm$0.373& 4.829$\pm$0.520& {\bf{+39.4}}&  5.483$\pm$0.474& 5.444$\pm$0.592& {\bf{+24.7}} \\
\hline 
M$_{1,2}$& 0.335$\pm$0.034& 0.293$\pm$0.119& {\bf{+248.0}}& 0.319$\pm$0.036& 0.336$\pm$0.123& {\bf{+236.2}}&  0.396$\pm$0.082& 0.404$\pm$0.246& {\bf{+198.7}}  \\
M$_{1,3}$& 0.840$\pm$0.089& 0.834$\pm$0.091& {\bf{+2.7}}&   0.774$\pm$0.079& 0.759$\pm$0.082& {\bf{+3.8}}&   0.708$\pm$0.197& 0.724$\pm$0.277& {\bf{+40.2}} \\
M$_{1,4}$& 0.362$\pm$0.048& 0.367$\pm$0.052& {\bf{+9.3}}& 0.399$\pm$0.041& 0.393$\pm$0.049& {\bf{+18.6}}&  0.415$\pm$0.081& 0.436$\pm$0.084& {\bf{+4.2}}\\
M$_{2,3}$& 2.438$\pm$0.292& 2.472$\pm$0.518& {\bf{+77.1}}&  2.218$\pm$0.309& 2.359$\pm$0.410& {\bf{+32.8}}&  1.784$\pm$0.366& 1.710$\pm$0.410& {\bf{+12.0}} \\
M$_{2,4}$& 1.070$\pm$0.138& 1.102$\pm$0.236& {\bf{+69.9}}&   1.220$\pm$0.138& 1.232$\pm$0.200& {\bf{+44.8}}&  1.068$\pm$0.165& 1.050$\pm$0.215& {\bf{+29.8}} \\
M$_{3,4}$& 0.447$\pm$0.027& 0.454$\pm$0.037& {\bf{+38.6}}&  0.524$\pm$0.034& 0.518$\pm$0.048& {\bf{+42.5}}&   0.608$\pm$0.064& 0.596$\pm$0.071& {\bf{+10.9}} \\
\hline 
b$_{1}$& 0.946$\pm$0.113& 1.003$\pm$0.114& {\bf{+1.7}}& 1.046$\pm$0.129& 1.035$\pm$0.129& {\bf{+0.1}}&  1.200$\pm$0.162& 1.240$\pm$0.168& {\bf{+3.8}}  \\
b$_{2}$& 1.878$\pm$0.148& 1.946$\pm$0.251& {\bf{+69.9}}& 2.119$\pm$0.200& 2.119$\pm$0.252& {\bf{+26.1}}&  2.140$\pm$0.220& 2.078$\pm$0.286& {\bf{+30.0}} \\
b$_{3}$& 1.050$\pm$0.103& 1.194$\pm$0.0& {\bf{--}}&   1.357$\pm$0.117& 1.357$\pm$0.113& {\bf{-3.6}}&  1.489$\pm$0.135& 1.508$\pm$0.140& {\bf{+3.5}}\\
b$_{4}$& 1.786$\pm$0.112& 1.818$\pm$0.139& {\bf{+24.1}}& 1.868$\pm$0.134& 1.946$\pm$0.146& {\bf{+8.9}}&  1.997$\pm$0.140& 2.002$\pm$0.161& {\bf{+15.1}} \\

\hline  

\end{tabular}
\end{center}
\end{table*}


\begin{table*}      
\caption{Same as Table~\ref{tab:gains1} for an integration range $0.1 < k[Mpc/h] < 0.3$ in the W4 field.}    
\begin{center}
\label{tab:gains3}   
\begin{tabular}{c | ccc | ccc | ccc}   
\hline &     \multicolumn{3}{ |c| }{$0.6<z<0.75$}& \multicolumn{3}{ |c| }{$0.75<z<0.9$}& \multicolumn{3}{ |c| }{$0.9<z<1.1$} \\
\hline  Param. &   MTOE& FKP& Gain($\%$)&  MTOE& FKP& Gain($\%$)& MTOE& FKP& Gain($\%$)\\
\hline 
M1& 1.538$\pm$0.199& 1.553$\pm$0.172& {\bf{-13.9}}&  1.946$\pm$0.222& 1.964$\pm$0.220& {\bf{-0.7}}&  2.386$\pm$0.364& 2.443$\pm$0.334& {\bf{-8.2}} \\
M2& 3.256$\pm$0.590& 2.870$\pm$0.863& {\bf{+46.2}}&  5.377$\pm$0.881& 5.248$\pm$1.065& {\bf{+20.9}}&   5.650$\pm$0.983& 5.402$\pm$1.335& {\bf{+35.8}}  \\
M3& 2.019$\pm$0.184& 1.958$\pm$0.181& {\bf{-1.8}}&  2.454$\pm$0.203& 2.414$\pm$0.201& {\bf{-1.2}}&  5.650$\pm$0.983& 5.402$\pm$1.335& {\bf{+35.8}} \\
M4& 3.134$\pm$0.451& 3.078$\pm$0.490& {\bf{+8.5}}&  4.560$\pm$0.534& 4.783$\pm$0.637& {\bf{+19.3}}& 5.319$\pm$0.546& 5.135$\pm$0.663& {\bf{+21.5}}\\
\hline 
M$_{1,2}$& 0.489$\pm$0.083& 0.474$\pm$0.240& {\bf{+186.7}}&  0.363$\pm$0.059& 0.359$\pm$0.074& {\bf{+24.0}}&  0.405$\pm$0.062& 0.416$\pm$0.102& {\bf{+65.1}}  \\
M$_{1,3}$& 0.769$\pm$0.098& 0.736$\pm$0.085& {\bf{-13.5}}&  0.800$\pm$0.089& 0.793$\pm$0.087& {\bf{-3.2}}&  0.693$\pm$0.101& 0.729$\pm$0.098& {\bf{-3.0}}\\
M$_{1,4}$& 0.476$\pm$0.078& 0.476$\pm$0.077& {\bf{-2.2}}&  0.424$\pm$0.055& 0.425$\pm$0.063& {\bf{+13.8}}&  0.443$\pm$0.070& 0.459$\pm$0.072& {\bf{+2.5}}\\
M$_{2,3}$& 1.629$\pm$0.275& 1.566$\pm$0.417& {\bf{+51.6}}&  2.242$\pm$0.336& 2.224$\pm$0.386& {\bf{+14.9}}& 1.661$\pm$0.272& 1.562$\pm$0.355& {\bf{+30.4}} \\
M$_{2,4}$& 0.981$\pm$0.160& 0.951$\pm$0.254& {\bf{+58.9}}&  1.156$\pm$0.159& 1.160$\pm$0.203& {\bf{+27.4}}&  1.042$\pm$0.159& 1.102$\pm$0.215& {\bf{+34.8}} \\
M$_{3,4}$& 0.636$\pm$0.070& 0.644$\pm$0.091& {\bf{+30.0}}&  0.521$\pm$0.049& 0.515$\pm$0.063& {\bf{+30.1}}& 0.611$\pm$0.063& 0.640$\pm$0.069& {\bf{+8.6}}\\
\hline 
b$_{1}$& 0.977$\pm$0.134& 0.916$\pm$0.119& {\bf{-11.1}}&   1.086$\pm$0.132& 1.034$\pm$0.132& {\bf{-0.2}}&  1.251$\pm$0.162& 1.228$\pm$0.147& {\bf{-9.2}}  \\
b$_{2}$& 1.418$\pm$0.197& 1.490$\pm$0.274& {\bf{+39.4}}&   1.984$\pm$0.215& 1.974$\pm$0.255& {\bf{+18.8}}&  2.083$\pm$0.227& 2.099$\pm$0.299& {\bf{+31.8}}  \\
b$_{3}$& 1.181$\pm$0.119& 1.116$\pm$0.114& {\bf{-4.6}}&   1.194$\pm$0.123& 1.280$\pm$0.124& {\bf{+0.6}}&  1.490$\pm$0.128& 1.481$\pm$0.126& {\bf{-1.4}} \\
b$_{4}$& 1.478$\pm$0.158& 1.497$\pm$0.167& {\bf{+5.5}}&   1.769$\pm$0.158& 1.863$\pm$0.175& {\bf{+11.2}}& 2.001$\pm$0.152& 2.005$\pm$0.169& {\bf{+11.1}} \\

\hline  

\end{tabular}
\end{center}
\end{table*}

\begin{table*}      
\caption{Same as Table~\ref{tab:gains1} for an integration range $0.3 < k[Mpc/h] < 0.5$ in the W4 field.}    
\begin{center}
\label{tab:gains4}   
\begin{tabular}{c | ccc | ccc | ccc}   
\hline &     \multicolumn{3}{ |c| }{$0.6<z<0.75$}& \multicolumn{3}{ |c| }{$0.75<z<0.9$}& \multicolumn{3}{ |c| }{$0.9<z<1.1$} \\
\hline  Param. &   MTOE& FKP& Gain($\%$)&  MTOE& FKP& Gain($\%$)& MTOE& FKP& Gain($\%$)\\
\hline 
M1& 1.500$\pm$0.180& 1.591$\pm$0.240& {\bf{+33.5}}&    1.927$\pm$0.235& 1.877$\pm$0.278& {\bf{+18.0}}&       2.379$\pm$0.529& 2.462$\pm$0.641& {\bf{+21.3}} \\
M2& 3.096$\pm$0.914& 3.374$\pm$1.456& {\bf{+59.3}}&    5.258$\pm$1.096& 5.537$\pm$1.829& {\bf{+66.9}}&       5.599$\pm$1.258& 4.964$\pm$2.137& {\bf{+69.9}}   \\
M3& 1.956$\pm$0.174& 2.030$\pm$0.154& {\bf{-11.2}}&    2.386$\pm$0.191& 2.387$\pm$0.230& {\bf{+20.1}}&       3.398$\pm$0.367& 3.357$\pm$0.414& {\bf{+12.9}}  \\
M4& 3.119$\pm$0.601& 3.081$\pm$0.654& {\bf{+8.7}}&     4.510$\pm$0.589& 4.553$\pm$0.745& {\bf{+26.3}}&      5.140$\pm$0.682& 5.151$\pm$0.880& {\bf{+29.1}} \\
\hline 
M$_{1,2}$& 0.491$\pm$0.252& 0.465$\pm$0.368& {\bf{+46.0}}&  0.373$\pm$0.144& 0.211$\pm$0.175& {\bf{+21.3}}&  0.412$\pm$0.130& 0.432$\pm$0.350& {\bf{+168.5}}  \\
M$_{1,3}$& 0.725$\pm$0.108& 0.720$\pm$0.133& {\bf{+23.4}}&  0.813$\pm$0.109& 0.761$\pm$0.119& {\bf{+9.0}}&  0.693$\pm$0.155& 0.697$\pm$0.269& {\bf{+72.9}}\\
M$_{1,4}$& 0.464$\pm$0.118& 0.474$\pm$0.352& {\bf{+196.5}}&  0.413$\pm$0.067& 0.409$\pm$0.087& {\bf{+30.0}}&  0.436$\pm$0.106& 0.460$\pm$0.141& {\bf{+32.5}}\\
M$_{2,3}$& 1.515$\pm$0.426& 1.467$\pm$0.725& {\bf{+70.2}}&  2.153$\pm$0.437& 2.297$\pm$0.714& {\bf{+63.4}}&  1.707$\pm$0.529& 1.476$\pm$1.224& {\bf{+131.3}} \\
M$_{2,4}$& 0.990$\pm$0.298& 0.930$\pm$0.750& {\bf{+150.9}}&  1.177$\pm$0.203& 1.108$\pm$0.385& {\bf{+89.7}}&  1.079$\pm$0.240& 0.955$\pm$0.407& {\bf{+69.2}} \\
M$_{3,4}$& 0.612$\pm$0.130& 0.431$\pm$0.241& {\bf{+85.2}}&  0.513$\pm$0.061& 0.508$\pm$0.080& {\bf{+31.0}}&  0.635$\pm$0.084& 0.645$\pm$0.103& {\bf{+22.7}}\\
\hline 
b$_{1}$& 0.933$\pm$0.124& 0.940$\pm$0.138& {\bf{+10.8}}&   1.070$\pm$0.127& 1.035$\pm$0.144& {\bf{+12.6}}&   1.200$\pm$0.207& 1.219$\pm$0.238& {\bf{+14.9}}  \\
b$_{2}$& 1.484$\pm$0.269& 1.628$\pm$0.449& {\bf{+66.5}}&   2.019$\pm$0.254& 2.026$\pm$0.420& {\bf{+65.2}}&   2.083$\pm$0.287& 1.973$\pm$0.473& {\bf{+64.5}} \\
b$_{3}$& 1.142$\pm$0.110& 1.162$\pm$0.115& {\bf{+4.6}}&   1.319$\pm$0.121& 1.293$\pm$0.128& {\bf{+5.6}}&   1.475$\pm$0.149& 1.466$\pm$0.155& {\bf{+3.7}}\\
b$_{4}$& 1.474$\pm$0.188& 1.496$\pm$0.214& {\bf{+13.8}}&   1.823$\pm$0.169& 1.786$\pm$0.202& {\bf{+19.6}}&   1.949$\pm$0.179& 1.850$\pm$0.217& {\bf{+20.8}}  \\

\hline  

\end{tabular}
\end{center}
\end{table*}



\label{lastpage}

\end{document}